\documentclass[authoryear]{article}
\usepackage[english]{babel}
\usepackage{fancyhdr} 
\usepackage[pdftex]{graphicx} 
\usepackage{url} 
\usepackage{caption} 
\usepackage{subcaption} 
\usepackage{natbib} 
\usepackage{multirow} 
\usepackage{float} 
\usepackage{amssymb}
\usepackage[intlimits]{amsmath} 
\usepackage{amsthm} 
\usepackage[german]{nomencl}
\usepackage{booktabs} 
\usepackage{geometry}
\usepackage{enumitem}
\makenomenclature
\usepackage[pdfborder={000},colorlinks=true,linkcolor=blue,citecolor=blue]{hyperref}
\usepackage{listings}
\usepackage[all]{xy}
\usepackage{color}

\definecolor{b}{rgb}{0,0,.8}	
\definecolor{g}{rgb}{0,.6,0}	
\definecolor{n}{rgb}{0,0,0}	
\definecolor{h}{rgb}{0.4,0.2,0.2}	
\definecolor{v}{rgb}{0.2,0.6,0}

\newcommand{\E}{{\mathbb E}}

\newcommand{\G}{{\mathbb G}}

\newcommand{\N}{{\mathbb N}}

\newcommand{\R}{{\mathbb R}}
\newcommand{\T}{{\mathbb T}}

\newcommand{\V}{{\mathbb V}}

\newcommand{\Z}{{\mathbb Z}}

\newcommand{\CC}{{\mathcal{C}}}

\newcommand{\II}{{\mathcal{I}}}

\newcommand{\KK}{{\mathcal{K}}}
\newcommand{\LL}{{\mathcal{L}}}

\newcommand{\OO}{{\mathcal{O}}}
\newcommand{\PP}{{\mathcal{P}}}

\newcommand{\RR}{{\mathcal{R}}}

\newcommand{\bsw}{\boldsymbol w}

\newcommand{\bsB}{\boldsymbol B}

\newcommand{\bsG}{\boldsymbol G}

\newcommand{\bsW}{\boldsymbol W}
\newcommand{\bsX}{\boldsymbol X}
\newcommand{\bsY}{\boldsymbol Y}
\newcommand{\bsZ}{\boldsymbol Z}
\newcommand{\bsone}{\boldsymbol 1}

\newcommand{\bsnull}{\boldsymbol 0}

\newcommand{\bsalpha}{\boldsymbol \alpha}

\newcommand{\bsmu}{\boldsymbol \mu}
\newcommand{\bseps}{\boldsymbol \varepsilon}
\newcommand{\bstheta}{\boldsymbol \theta}
\newcommand{\bssigma}{\boldsymbol \sigma}
\newcommand{\bszeta}{\boldsymbol \zeta}

\newcommand{\bsGamma}{\boldsymbol \Gamma}

\newcommand{\eps}{{\varepsilon}}

\DeclareMathOperator*{\argmin}{arg\,min}

\DeclareMathOperator{\modulo}{mod}
\DeclareMathOperator{\modulodiv}{div}

\DeclareMathOperator{\var}{\V ar}

\newcommand{\ov}\overline
\newcommand{\what}{\widehat}
\newcommand{\wtilde}{\widetilde}

\newcommand{\rig}\right
\newcommand{\lef}\left
\newcommand{\nf}\normalfont

\newcommand{\MAE}{\text{MAE}} 
\newcommand{\MMAE}{\text{MMAE}} 
\newcommand{\bsMAE}{\text{\textbf{MAE}}} 
\newcommand{\bsMMAE}{\text{\textbf{MMAE}}}

\begin{document}

\title{Efficient Modeling and Forecasting \\ of Electricity Spot Prices}

\author{Florian~Ziel, Rick~Steinert, Sven~Husmann}



\maketitle

\begin{abstract}
The increasing importance of renewable energy, especially solar and wind power, 
has led to new forces in the formation of electricity prices. Hence, this paper introduces an econometric model for the hourly time series of electricity prices of the European Power Exchange (EPEX) 
which incorporates specific features like renewable energy. The model consists of several sophisticated and established
approaches and can be regarded as a periodic VAR-TARCH with wind power, solar power, and load as influences on the time series. It is able to map the distinct and well-known features of electricity prices in Germany. An efficient iteratively reweighted lasso
approach is used for the estimation. Moreover, it is shown that several existing models are outperformed by  the procedure developed in this paper.
\end{abstract}

\section{Introduction} \label{Introduction}
With the ongoing liberalization of electricity markets over the past decades,
the volume of electricity traded via exchanges has greatly increased. This in turn led to an increasing transparency of the price for electricity. 
Due to the substantial dependence of companies and private households on this price, modeling electricity prices 
has become one of the cornerstones of research into the energy markets. But such modeling turns out to be at 
the edge of many disciplines in research. For instance, the analysis of the underlying trade mechanisms can
be allocated to economics. But the energy production itself is a process which can be related to engineering and the 
rules governing the exchange of energy, especially renewable energy production, are determined by law and 
politics. Hence, modeling electricity prices can be a complex issue. This is also reflected in the time series,
where many unusual but already stylized facts can be observed. \par 
The model proposed in our paper tackles this complexity in several ways. 
Its distinctive features compared to the existing literature can be boiled down to seven key facts. Our approach: 1. Models the electricity price without any data manipulation, 2. Incorporates every established stylized fact of electricity prices,
3. Provides insights for the structure of the leverage effect, 4. Proves the effect of wind and solar power on price, 
5. Accounts for specific holiday effects and daylight saving time effects in the wind and solar generation, 
6. Does not need any future information to provide accurate forecasts. Finally, 7, it uses efficient and rapid state of the art estimation techniques.   

\par
We will fit our proposed model to the hourly electricity price of the European Power Exchange (EPEX) for the period of 28.09.2010 up to 01.05.2014. The data sets are obtained 
from the EPEX at www.epexspot.com for the hourly day-ahead spot price data of Germany/Austria, from the European Network of Transmission System Operators for Electricity at www.entsoe.eu for the hourly load data of Germany, and from the Transparency Platform of the European Energy Exchange (EEX) at www.transparency.eex.com for the hourly wind and solar power feed-in for Germany.\footnote{We remark that the load and renewable data for Austria is not included in the data as our main focus is the German energy market.}\par
Our paper is organized as follows. In Section \ref{Challenges in modeling electricity prices} 
we give an overview about the setting in which electricity exchange takes place and name the distinct challenges
occurring in modeling the time series. The subsequent section sets up our model, which aims to face those challenges. Section \ref{Estimation} presents the 
estimation procedures for our model. In Sections \ref{Results} and \ref{Forecasting} we fit the model to the hourly electricity price time series of the European Power Exchange and apply a comprehensive forecast study. The last section concludes by discussing our findings.

\section{Challenges in modeling electricity prices} \label{Challenges in modeling electricity prices}
Energy markets are a rapidly changing field of the economy. The liberalization of these markets and the subsequent development of the energy mix account for that fact. But as different countries worldwide face different preconditions, for instance politically or climatically, their energy markets tend to have a very heterogeneous structure. Hence, the findings for one country may not, or may only partly, be used for another country or region. \par 
In case of the German energy market, a substantial amount of the daily demand is traded via an exchange. Spot market trading takes place by continuous trading and auctions. Prices for this market can be obtained either from intraday trading or from day-ahead auction prices. The latter is represented by the EPEX spot auction price for Austria and Germany. EPEX is a member of the EEX group. Since 2008, 
the EEX has not prohibited negative prices \citep{keles2012comparison}. By considering all products of the exchange, there are currently 250 market participants at the EEX. Considering only the Austrian/German day-ahead auction of the EPEX results in 197 participating traders. 
\par 
The non-negligible amount of market participants is also an aftermath of liberal energy laws in Germany, especially when renewable energy is considered. The Erneuerbare-Energien-Gesetz (EEG) and its corresponding enactments are governmental regulations which embody this idea of liberalization. As a consequence, the supply of energy from renewables rose significantly within the past years. Incentives, like feed-in-tariffs for renewables granted by the government, catalyzed this development. But the growing amount of power plants for renewables had a direct influence on the price for electricity,  \citep{edenhofer2013economics}, as such energy can be produced at a cost of almost zero \citep{wurzburg2013renewable}. 
\par
Changes in the EEG in 2012 had direct impact on the marketing of renewable
energy.\footnote{Even more recent changes of the EEG in August 2014 are not considered within this article, as our dataset ends in May 2014.} 
The EEG allows for marketing the produced renewable energy not only to
TSOs or other market players but also directly at the exchange. According to \S 33g EEG, producers of such energy 
receive a market premium in addition to the market price for selling their electricity. Hence, also producers of 
renewable energy have an incentive to directly trade at the electricity exchange or to sell it via OTC business.
\par
Nevertheless, other regulations have also affected the market directly. For instance, transmission system operators (TSO) are obliged to sell their electricity only on the day-ahead or intraday spot market, if they decide to use an exchange.\footnote{According to \S 2 AusglMechV} Hence, data since the publication of this regulation in 2009 are of special interest.
\par
In the economic theory of competitive markets, the price of electricity should equal its marginal cost. 
As the feed-in of renewable energy with zero marginal cost replaces every other energy source with higher marginal cost, 
the price of electricity should decline \citep{keles2013combined}. This is known as the merit-order effect. Furthermore, 
the demand curve for electricity can be assumed to be inelastic \citep{sensfuss2008merit} as a 
certain amount of power is needed regardless of the price. This implies that modeling the impact
of renewables on the price of electricity is highly necessary, as those energy sources will always lead to a modified
marginal cost structure as the share of renewable energy sources is increasing. \par
Empirical evidence for the reduction of electricity prices caused by the emergence of renewable energy has been shown 
by many authors in the recent literature. For instance, \cite{woo2011impact} use a regression analysis for the Texas 
electricity price market to examine the effect of wind power generation. Another multivariate regression approach was 
applied to German and Austrian electricity prices by \cite{wurzburg2013renewable}, where, among other time series, wind 
and solar power were examined and also led to a reduction in price. \cite{huisman2013renewable} obtained equivalent results 
for the Nord Pool market by modeling energy supply and demand. \par
\begin{figure}[hbt!] 
\centering
 \includegraphics[width=0.95\textwidth]{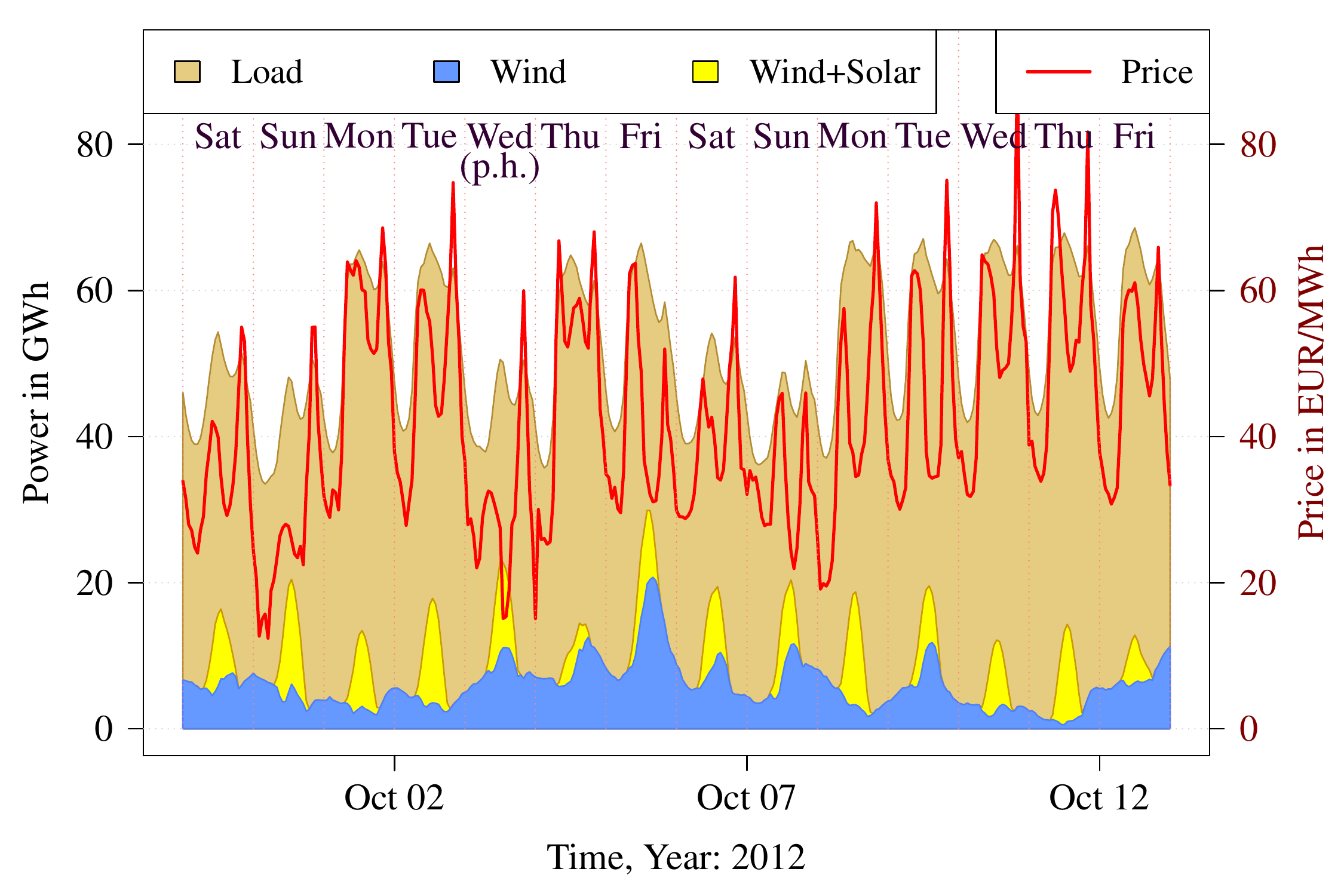}
 \caption{The hourly load, wind and solar power feed-in with the corresponding spot price. 
 The 3rd of October is a public holiday (German unity).}
 \label{fig_load-wind-solar}
\end{figure}

This relation is also illustrated in Figure \ref{fig_load-wind-solar}, which shows the price, load, and the patterns 
of solar power and wind power for two weeks of October 2012. Whenever the combined effect of wind and solar power is high, the line representing the price seems to decline considerably. \par

But the introduction of renewables not only leads to a price reduction effect, it can also increase the goodness-of-fit for modeling the time series of prices.
Concerning this, a comprehensive study was done by \cite{cruz2011effect}. They combined sophisticated models like Holt--Winters, ARIMA, and neural networks, with, e.g., wind power, and provided evidence for this inclusion's being beneficial to the price modeling. Also \cite{yan2013mid}, \cite{liebl2013modeling} and \cite{kristiansen2012forecasting} obtained competitive goodness-of-fit statistics by including wind power in their approaches. Using a simulation study for the dependencies of the EEX electricity price and the wind power generation \cite{keles2013combined} were able to show the advantages of including wind power.
\par
The price of electricity is also heavily dependent on the day of the week and also changes over a year, which last is primarily governed by the four seasons of the year \citep{weron2006modeling}. The reason for this is twofold. First, especially solar energy production depends by a law of nature on the period of sunshine: reduced, for instance, in winter. Second, the daily demand for electricity is dependent on the working days, e.g. whether industrial machines are running and require energy or not. 
An example of two observed months for illustrative purposes is given in Figure \ref{fig_load-wind-solar}. The light-brown shaded area represents the electric load pattern of the first two weeks of October in 2012. Comparing the price time series with the load time series indicates that both variables are positively related. As on weekends the load is usually lowered, weekends and weekdays\footnote{We use the term weekday to refer to the days from Monday to Friday} within the chosen period tend to exhibit different price patterns. This is also true for holidays, e.g. German unity day on the 3rd of October, where the price and the load are both lower than on other weekdays. A more detailed investigation of special days and phases of the day is done in Section \ref{Modeling electricity prices}. \par 
The consideration of these effects is also an important topic in the literature. The weekly dependence is usually modeled in time series analysis by incorporating the equivalent lagged value, e.g. lag 168 for hourly data. (e.g. in \cite{kristiansen2012forecasting}) Nevertheless, the consideration of special days or special phases of the day is done only rarely. \cite{cancelo2008forecasting} use a combined ARIMA model to forecast load and provide evidence for the inclusion of special days within the estimation. \cite{guthrie2007electricity} calculate a periodic autoregression for the half-hourly electricity price of the New Zealand Electricity Market. They divide the day into five different time periods and show that they differ from each other. \par 
Another appealing approach is to let the coefficients of the proposed model vary over time. For instance, \cite{karakatsani2008forecasting} present a comparative analysis of models with and without time varying parameters. They show for British half-hourly electricity prices that models with time varying parameters dominate other autoregressive approaches with constant parameters. A combined model for the inclusion of holidays and time-varying coefficients was introduced by \cite{koopman2007periodic}. They use a Reg-ARFIMA-GARCH model for the EEX, Powernext and AAX price data.
\par
Electricity is also distinct from most other commodities as it is not efficiently storable with the existing technology. This, therefore, adds another source of risk. \citep{knittel2005empirical} Nevertheless, to a certain extent it can be regarded as indirectly storable because some energy sources such as fuel or gas can be stored and therefore used equivalently to fulfill obligations from, e.g., derivatives. \citep{huisman2012electricity} \par
The combination of almost non-storability, inelastic demand, and the strong dependence on highly fluctuating energy supplies causes the price of electricity to be endowed with unique characteristics. These are known as the stylized facts of electricity prices.  \par
\begin{figure}[hbt!]
\centering
 \includegraphics[width=0.95\textwidth]{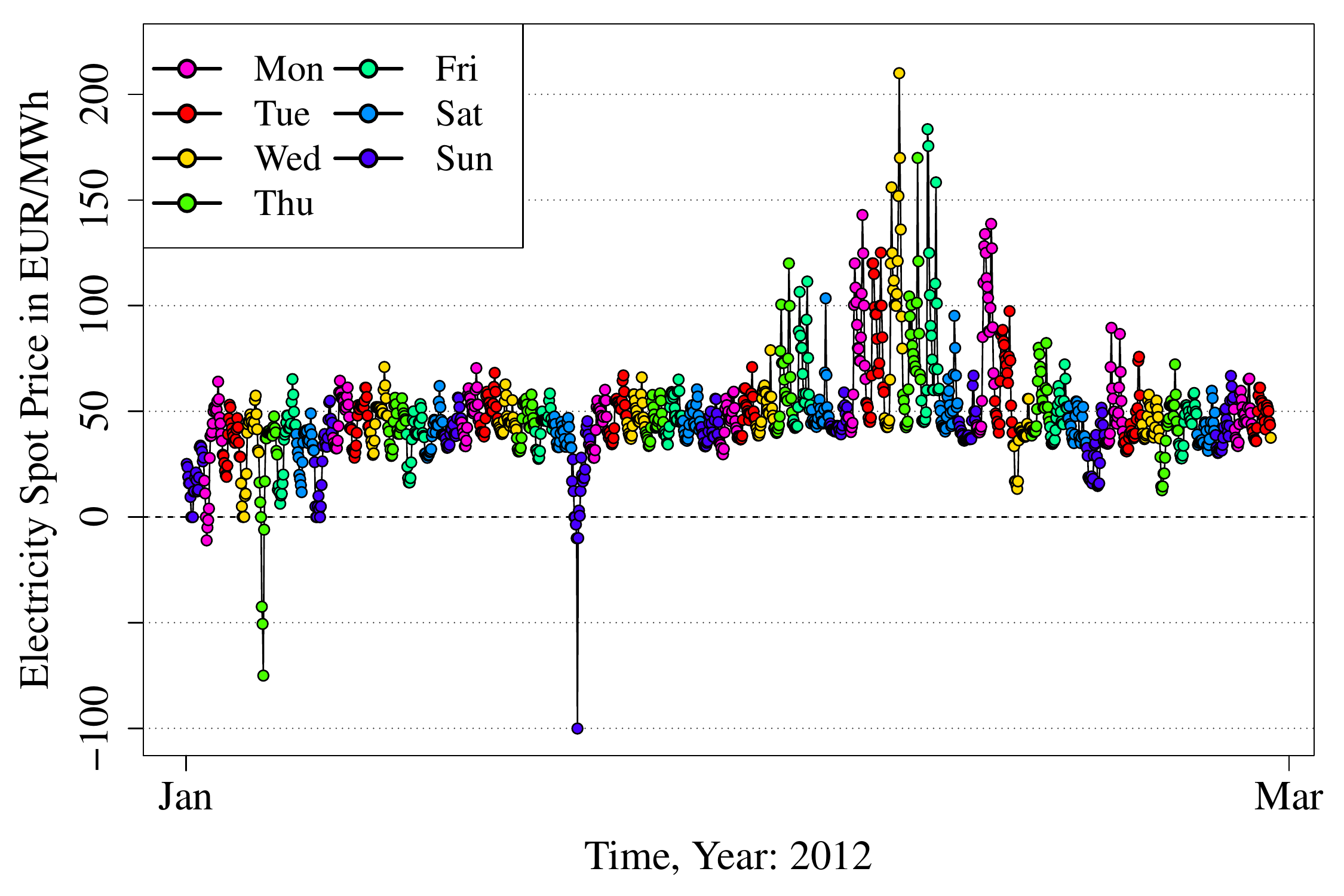}
 \caption{Hourly electricity spot price from 1st January to 1st March 2012.}
 \label{fig_price-sample}
\end{figure}

Most authors refer to three or even more characteristics. The most frequently mentioned are seasonality, mean-reversion, and high heteroscedastic volatility with extreme price spikes. (E.g. \cite{weron2006modeling} and \cite{eydeland2003energy}.) For illustrative purposes, the EPEX electricity price of the first two months of the year 2012 are depicted in Figure \ref{fig_price-sample}. The pattern shows high price spikes in both directions, which usually last for several hours. After the impact of such a shock, the price reverts to its usual daily level.\par 

Empirical consideration of negative prices in the German/Austrian electricity market is quite rare as they were not present until 2009. Moreover, modeling negative prices is also a difficult task, 
as some fitting and transformation approaches like logarithmic transformation are not possible. 
Hence, some authors simply cut off or shift the time series. Nevertheless, \cite{keles2012comparison} showed via simulation study for the EEX that the inclusion of negative prices yielded better results for their investigated models. Also \cite{fanone2013case} argue for the inclusion of negative prices. \par
Price spikes can be modeled in different ways. According to \cite{christensen2012forecasting} there are mainly three different common approaches. These are specific autoregressive, Markov-regime-switching, and jump diffusion models. Other authors take advantage of a combination of these models, e.g. \cite{bordignon2013combining} and \cite{escribano2011modelling}.
Besides these obvious characteristics of electricity prices, the recent literature has argued for an appreciation of 
the leverage effect, which is well-known in financial economics. A leverage effect accounts for the variance of a 
time series's reacting asymmetrically to negative and positive price shocks. \citep{black1976leverage} The standard 
leverage effect has higher variance for negative than for positive price shocks. By analyzing hourly California 
electricity prices \cite{knittel2005empirical} were able to detect an inverse leverage effect for the time series. 
Their paper showed that the volatility responses were more intense to positive than to negative shocks. This was later 
acknowledged for instance by \cite{bowden2008short} and \cite{liu2013applying}. \cite{cifter2013forecasting}, however, 
provided evidence for the standard leverage effect by examining the daily returns of the Nord Pool electricity price market. \par
As a matter of fact, modeling electricity prices is a complex issue. 
In addition to classic autoregression approaches, many researchers have applied methods 
from related scientific fields, e.g. heuristic or a combination of autoregression and other techniques. 
As a detailed overview of all those approaches would be outside the scope of this paper, the interested reader
is referred to the works of \cite{weron2006modeling} or \cite{aggarwal2009electricity} who provide a comprehensive overview
of the recent literature. \par
In the following section we will address the challenges which occur when electricity prices are examined. We introduce our own model,
which is tailor-made for the hourly EPEX electricity price data.
It will be shown in detail how the challenges influence our approach and how they are tackled. 
We will refer to this model as periodic
VAR-TARCH. It includes hourly data of the electricity load, and of wind and solar power feed-in. The model is capable of forecasting the 
price itself and also its random dependent variables.

\section{Model for electricity prices} \label{Modeling electricity prices}
Let $\bsY_t$ be the considered multivariate process with dimension $d=3$ . So
$\bsY_t = \left(Y_t^{\PP}, Y_t^{\LL}, Y_t^{\RR} \right)$ is a vector of the price of electricity, the load,
and the renewable power feed-in at time $t$. Note that $Y_t^{\RR}$ is the 
sum of the wind and solar power feed-in. Let $\II = \{\PP, \LL, \RR \}$ be the corresponding index set for the three 
univariate processes.

The basic model is almost equivalent to a simple vector autoregressive model.
It is given by the autoregressive model
\begin{equation}
 Y_t^i = \mu^i(t) + \sum_{j \in \II} \sum_{k\in I_{i,j}} \phi^{i,j}_k(t) Y_{t-k}^j+ \eps^i_t
 \label{eq_main_model}
\end{equation}
for $i\in \II$, where $\bsmu(t) = (\mu^\PP(t), \mu^\LL(t), \mu^\RR(t))$ is a trend component, 
$\bseps_t = \left(\eps^\PP_t, \eps^\LL_t, \eps^\RR_t \right)$ 
are noise terms with zero mean and time dependent covariance matrix $\Sigma_t$, and they are assumed to be independent.
Note that the autoregressive parameters $\phi^{i,j}_k(t)$ may also depend on time, such as the trend $\mu(t)$ and the
matrix $\Sigma_t$. 
The considered lags are given by the index sets $I_{i,j}$. 
These index sets are crucial and determine
the autoregressive dependency structure within $\bsY_t$. 
Here it is important to know 
that the estimation method described below contains an automatic model selection procedure.
Hence, there is always a trade-off in the pre-selection of index sets. Giving the model  
large index sets $I_{i,j}$ can lead to better models as there is more possible information provided. But on the other side
this increases the computation time and the probability of modeling noise. Thus, especially the index sets
$I_{i,j}$ have to be chosen carefully.

The index sets $I_{i,j}$ we used are given in Table \ref{tab_lags}.
\begin{table}[tbh]
\centering
\begin{tabular}{p{2.8cm}|p{10cm}}
  \hline
 Index sets & contained Lags  \\ 
  \hline \hline
$I_{\PP, \PP}$, $I_{\LL,\LL}$, $I_{\PP, \LL}$ & 1,2,$\ldots$,361,504,505,672,673,840,841,1008,1009\\ \hline 
$I_{\PP, \RR}$, $I_{\LL, \RR}$ & 1,2,$\ldots$,49 \\  \hline
$I_{\RR, \PP}$, $I_{\RR, \LL}, I_{\LL, \PP}$ & - \\  \hline
$I_{\RR, \RR}$ & 1,2,$\ldots$,361 \\ 
\end{tabular}
\caption{Considered lags of the index sets $L_{i,j}$ for $i,j\in \II$. }
\label{tab_lags}
\end{table}
On first view the chosen lags in $I_{i,j}$ seem more or less random, 
albeit every lag is chosen by employing statistical facts of the data.

For most index sets $I_{i,j}$ the first lags $1,2, \ldots 361 $ simply model the linear dependence of the past 361 hours, 
so 2 weeks plus a day and an hour. For $I_{\PP, \RR}$, $I_{\LL, \RR}$ we chose a dependence of the past 49 hours.
But we also added some larger lags like $504$ and $505$ to $I_{\PP, \PP}$, $I_{\LL,\LL}$, and $I_{\PP, \LL}$ which account for 
a possible dependence on the price of the past weeks. 

These lags allow a structure in the process
that is similar to a multiplicative seasonal AR model. 
For example, the multiplicative structure of the lag
polynomials in a seasonal AR(1)$\times$(2)$_{168}$ is given by 
$(1-a_1B)(1-a_{168}B^{168}- a_{336}B^{336}) = 
1- a_1B - a_{168}B^{168} + a_{1}a_{168}B^{169}  + a_{336}B^{336} + a_{1}a_{336}B^{337} $ with $B$ as backshift- resp. lag operator. 

Thus it contains lag 1, 168, 169, 336, 337. Using this evauation of the lag-polynomial for 
higher order seasonal AR processes, we receive that lags like 504, 505, 672, 673, 840, 841 or 1008,1009
should be contained in the model.
Such seasonal lags were also considered by different authors in the past, such as \cite{liu2013applying}.

As mentioned above, we could also increase index sets, e.g. replacing $I_{\PP,\PP}$ by $\{1,\ldots,1009\}$.
However, we noticed during our study that many parameters between $(k-1)168+1$ and $k168$ are zero for large $k$, i.e. $k\geq3$.
Another simple but useful approach for selecting appropriate index sets is based on
evaluating the sample partial autocorrelation function of $\bsY_t$; the lags that correspond to 
large absolute values should be included. Indeed this approach leads to very similar index sets as we use, especially
we observe large PACF values at lags $504$ and $505$ as discussed above.
The pre-selection of lags therefore provides some flexibility to the applicant, as lags, which are just assumed to be significant can be implemented very easily and will still undergo a statistical reliable lag selection process.

Note that we choose $I_{\PP, \RR}$ and $I_{\LL, \RR}$ as being small, as we assume that there is no impact 
of renewable energy feed-in to the price to last for more than two days.
Moreover, the renewable energy feed-in does not have any weekly seasonal structure,
only a daily and a yearly one. Thus, $I_{\RR,\RR}$ does not contain higher order weekly based lags like $504$ and $505$ as e.g. $I_{\PP,\PP}$.

The trend component $\bsmu(t)$ is modeled by a linear combination of $M(\mu^i)$ basis functions plus a linear trend, so we have
\begin{equation}
 \mu^i(t) = \mu^i_0 + \mu_{\text{lin.}}^i t + \sum_{j=1}^{M(\mu^i)} \mu^i_j \wtilde{B}^{\mu^i}_j(t)
\label{eq_mean}
 \end{equation}
for $i\in \II$ with parameters $\mu^i_j$ and basis functions $\wtilde{B}^{\mu^i}_j(t)$.
The structure of the used basis functions is inspired by periodic basis functions. 
This enables modeling recurring events like specific days of the week. Nevertheless, as some events like holidays are not periodically, we are allowing $\wtilde{B}^{\mu^i}_j(t)$ to be not strictly periodic.
As basic concept for the construction of $\wtilde{B}^{\mu^i}_j(t)$
we consider periodic B-splines as used in \cite{harvey1993forecasting}.
However, this approach using local B-splines was mainly chosen to increase the flexibility and the interpretability of the model. We therefore want to point out, that other, e.g. Fourier approaches, might lead to estimation results of comparable quality.


The chosen B-splines are denoted as $B_{T,d_\KK}(t) =B_{T,d_\KK}(t; \KK(d_\KK,T,D), D)$, 
where $D$ represents the degree of the spline, 
$\KK = \KK(d_\KK,T,D) = \{k_{0},\ldots, k_{D+1}\}$
the set of equidistant knots $ k_{0} <  \cdots < k_{D+1} $, that
have equal distance $d_\KK$ and are centered around $T$.
%
Hence, we have $k_{0}= T - d_\KK \frac{D+1}{2}$, $k_{D+1} = T + d_\KK \frac{D+1}{2}$ and
if we select an odd degree $D$ we get $k_{\frac{D+1}{2}} = T$.
The distance $d_\KK= k_1-k_{0}$ 
is important for applications as it accounts for the degree of approximation of the time dependent effects
and affects the amounts of parameter in the model, the smaller $d_\KK$ the better the corresponding approximation, but the larger
the considered parameter space.
For $D$ we choose $D=3$, which provides the popular cubic 
splines which are twice continuously differentiable.


For mirroring the movement of the electricity price data 
we want that $\mu^i(t)$ smoothly
changes over the days and during the week as well. So, for example, the coefficient
for Sunday at 8 am is different to the coefficients at Sunday 9 am or Monday 8 am, but it is equal for every Sunday at 8 am.

\begin{figure}[hbt!]
\centering
\includegraphics[width=0.95\textwidth]{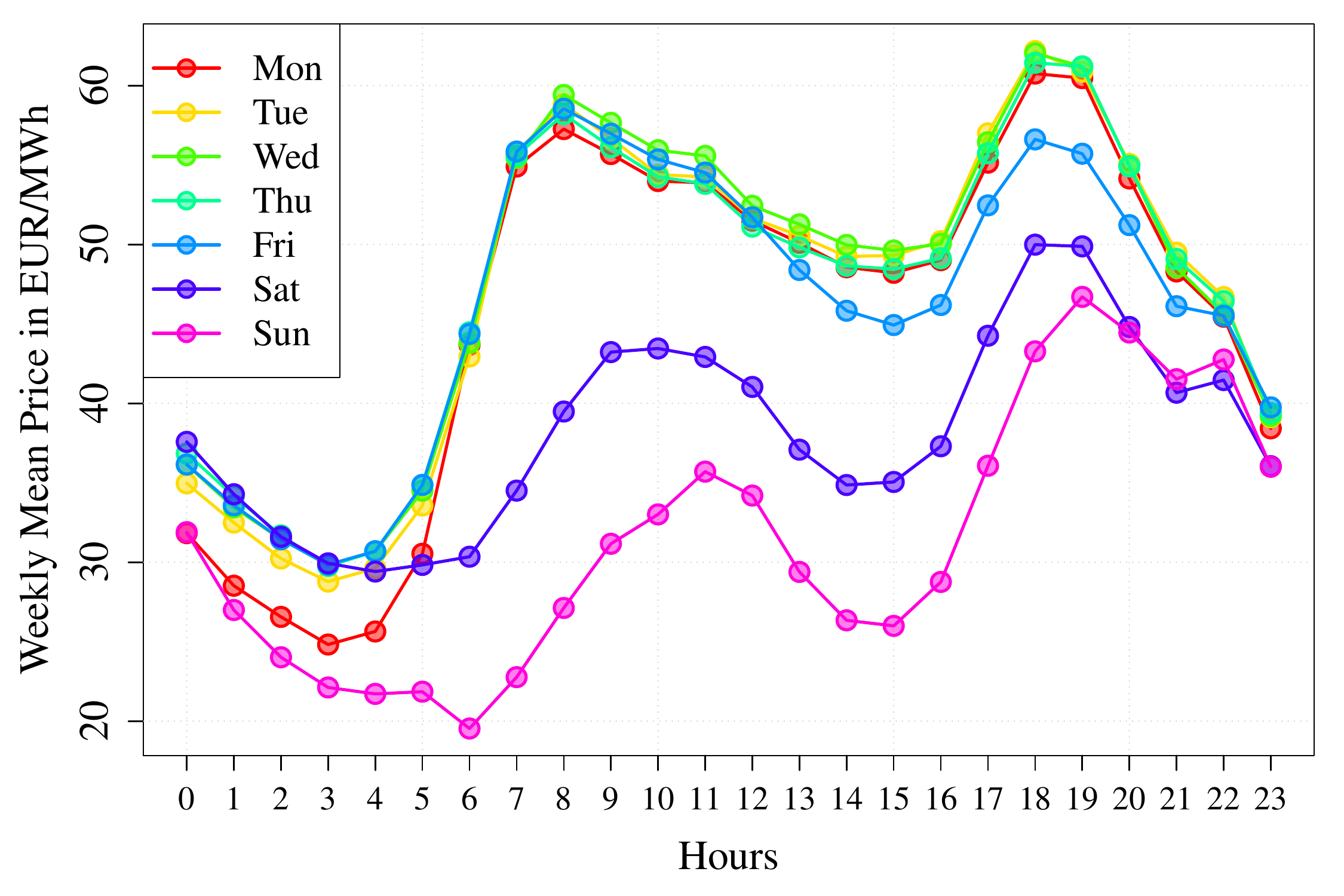}
 \caption{Average price in EUR/MWh for every hour, separated by every day of the week.}
 \label{fig_weekly_mean}
\end{figure}

However, we also observe that the mean behavior of at least some days is very similar to that of other days, see Figure \ref{fig_weekly_mean}. 
This finding is embedded in our model by proposing the assumption of some days having equal 
hourly mean components, which in turn reduces the number of parameters in the model.
For example, the mean at Tuesday 8 am is assumed to be the same as on a Wednesday 8 am. 
Finally we consider five groups of parameter restrictions: Saturday, Sunday, Monday from 0 am to 12 am, Friday from 12 am to 12 pm 
and the core week from Monday 12 am to Friday 12 am. 
The Monday morning and Friday afternoon and evening has to be considered to get an appropriate phase-in and
phase-out of the weekend. Friday, represented as the cyan colored line, 
departs from the typical structure of the other days (excluding the weekend) approximately at 12 am. It is almost the same for Monday, 
with the difference that Monday needs 7 to 8 hours to reach an equivalent structure comparable to the other weekdays. 
In order to maintain equidistant phase-in and phase-out periods, we set both periods to 12 hours.

Furthermore, there are significant impacts of public holidays, as mentioned by several authors. 
We consider every German public holiday as a Sunday. In addition to that there are some regional public holidays in Germany which are celebrated only by some areas. Every regional public holiday that concerns at least 25\% of the population 
and is not on a Sunday, is treated as a Saturday as well. Moreover, we consider Christmas Eve and New Years Eve as regional public holidays. 
Consequently we take the 12 hours after a (regional) public holiday as Monday-am and
the 12 hours before every (regional) public holiday as Friday-pm. 
Of course, this applies only if the day itself is not a Sunday, Saturday or (regional) public holiday.
An overview of the explained day-grouping is given in Table \ref{tab_days}. 
In addition, the Table provides the number of unique hours within each group. Unique refers to an hour having one unique parameter. For instance, in case of the group \emph{normal}, 
there are only 24 unique hours, even though its contained days, from Monday lunch time to Friday lunch time, has 96 hours in total. This approach accounts for the mean of hours of different days within this group being very similar.

\begin{table}[tbh]
\centering
\begin{tabular}{r|c|p{3.5cm}}
  \hline
 \multirow{2}{*}{group} &  \multirow{2}{*}{contained days}  &  hours with\\
 &&different parameters \\
  \hline
 \emph{full off} & Sundays and public holidays   & $24$ \\
 \emph{semi off} & Saturdays and regional public holiday if not \emph{full off}  & $24$  \\
 \emph{phase in} & 12 hours before \emph{full} or \emph{semi off} if not \emph{full} or \emph{semi off} & $12$ \\
 \emph{phase out} & 12 hours after \emph{full} or \emph{semi off} if not \emph{full} or \emph{semi off} & $12$ \\
 \emph{normal} & others & $24$ \\
  \end{tabular}
\caption{Group structure of the days.}
\label{tab_days}
\end{table}

To employ our model we denote the corresponding five groups of Table \ref{tab_days} as 
 $\G = \{1, \ldots, 5\}$ and create a time set $\T_{g}$. Every group $g$ is an element of $\G$, namely $g\in \G$. 
 The combined time set $\T_{g}$ for all groups contains the integer value of the first hour of every group and day. 
 The subset for the first group $\T_{1}$, for instance, contains only those integer values of hours, which lay inside the 
 group \textit{full off} and represent the hour 0 am of this day. For the group \textit{phase in} and \textit{normal} 
 however, the subsets $\T_{3}$ and $\T_{5}$
 contain the values of the hour 12 am for each day. The differences of the elements of $\T_{g}$ are therefore usually 12,24 or 168,
 but can also be a different number, when a public holiday occurs. An example for such a set $\T_{g}$ could
 therefore be $\T_{1}=\lbrace\ldots,0,168,336,504,528,672,\ldots\rbrace$, where at hour 528 (Monday) a holiday occurred. \par
Within each group $g$ we set a specific number of basis functions. This number is related to the unique
hours every group has. We determine the vector $N_\G$ as $N_\G = (d_\KK^{\text{weekly}})^{-1} (24,24,12,12,24)'$.
Please notice that we divide the amount of unique hours by $d_\KK^{\text{weekly}}=4$, so that for instance the group 
\textit{full off} contains $24/4=6$ different basis functions. 
In order to label these basis functions distinctively, we introduce the vector
$C_\G = (d_\KK^{\text{weekly}})^{-1} (0,24,48,60,72,96-d_\KK^{\text{weekly}})' $, which contains the cumulative sum of 
the hours in $N_\G$ as entries.
The subtraction of $d_\KK^{\text{weekly}}$ in
the last element of $C_\G$ is to avoid the singularity issue that occurs as the sum of all basis 
functions is constant. Then we define the function $G(\iota,C_\G) = g$ for $ C_\G(g) < \iota \leq C_\G(g+1)$. The index 
variable $\iota$ is integer valued, starts from 1 and contains in our case every natural number up to $(96-4)/4=23$. Hence, 
the function $G(\iota,C_\G)$ matches every $\iota$, which represents the label of a basis function, to its specific group.

\begin{figure}[hbt!]
\centering
 \includegraphics[width=0.95\textwidth]{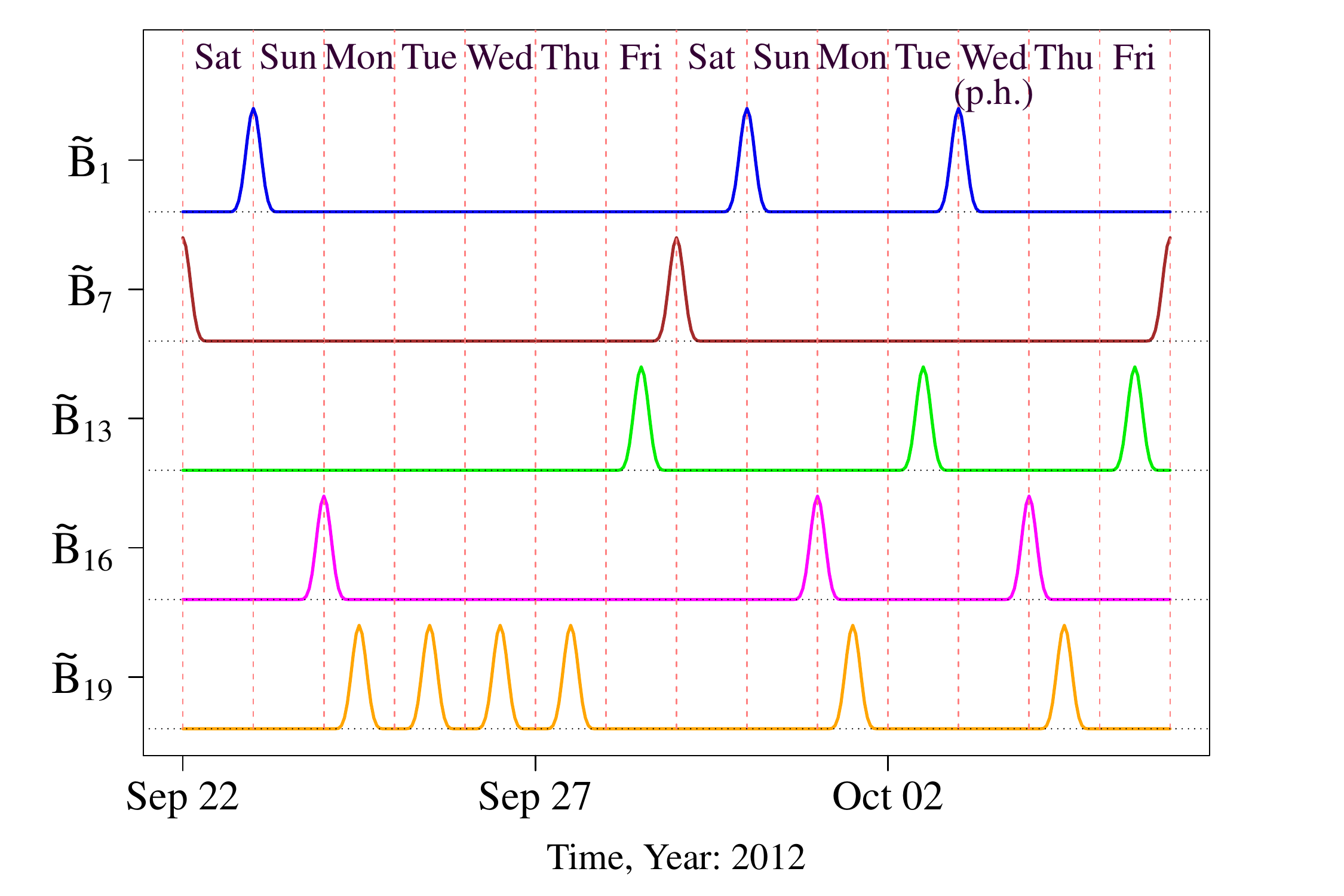}
 \caption{First basis function $\wtilde{B}^{\mu^i}_j$ for $i\in \{\PP, \LL\}$ and each group $g\in \G$ from 22.09.2012 to 05.10.2012.}
 \label{fig_basis}
\end{figure}

Moreover, we have to introduce the set $\CC = \{C_\G(g)+1 | g\in \G \}$ which will contain
the index of the first basis function of each group. In our case this turns out to be $\CC = \{1,7,13,16,19\}$.
The resulting basis functions $\wtilde{B}^{\mu^i}_j(t)$ can then be constructed as follows:
\begin{equation} \label{b-spline first}
 \wtilde{B}^{\mu^i}_j(t) = 
  \sum_{k \in \T_{G(j,C_\G) } } B_{k,d_\KK^{\text{weekly}}}(t), 
  \end{equation}
 for $j\in \CC$. The other basis functions can be obtained by shifting these basis functions.
 For $j \in \{ C_\G(g) +2, \ldots, C_\G(g+1)\}$ they are iteratively defined 
 as $\wtilde{B}^{\mu^i}_j(t)  = \wtilde{B}^{\mu^i}_{j-1}(t-d_\KK^{\text{weekly}}) $.
  These basis functions apply for the price and the load data, so $i\in \{\PP, \LL\}$. 
  Figure \ref{fig_basis} displays a practical application of equation (\ref{b-spline first}) 
  to the time from 22.09.2012 to 05.10.2012. It shows the first basis function of each of the five groups.
  The model automatically detects that the 03.10.2012 is a public holiday and therefore breaks the usual periodicity
  for some of the groups, whereas the other groups remain stable. The group \textit{normal} for instance leaves out 
  the basis functions for Tuesday and Wednesday, even though they would usually be considered within this group. 
  However, the group \textit{full off} does now include the Wednesday as public holiday and therefore gets a basis function for this day.

Furthermore, we can observe an annual pattern within the data, most distinctively for the load and solar data. 
Hence, we added B-splines with a yearly periodicity of $S_{\text{annual}}=24\times365.24 = 8765.76$. 
As we assume that the meteorological processes
have a smoother change over the year we added only $\eta^{\text{annual}}_{\RR} = 6$ basis functions, whereas we 
take $\eta^{\text{annual}}_{i} = 12$ basis functions for the price and load, so $i\in \{\PP, \LL \}$.
This leads to   $d_{\KK,\RR}^{\text{annual}}  = 1460.96$ and $d_{\KK,i}^{\text{annual}} = 730.48$. 
For the wind and solar components we assumed no weekly structure, only a daily and a yearly one. 
Therefore we construct basis function for the daily structure with a periodicity of $S_{\text{daily}} = 24$ and 
annual basis functions. We choose $d^{\text{daily}}_{\KK}=4$, 
thus there are $\eta^{\text{daily}} = 6$ basis functions added.
We also take into account interactions between the yearly and daily components, i.e. how the daily cycle behavior changes over the year. 
This seems important as, for instance, the length of the sunshine period of the day changes over 
the year and should have a strong influence. 
This is modeled by a multiplication of both components. 

In detail the additional basis functions for the price and load are given by 
$$ \wtilde{B}^{\mu^i}_{j}(t) = 
  \sum_{k \in   \Z } B_{k S_{\text{annual}} , d_{\KK,i}^{\text{annual}}}(t)  $$
for $j = (d_\KK^{\text{weekly}})^{-1} 96$ and $i\in \{\PP, \LL\}$.
The other basis functions can be obtained by shifting these basis functions.
 For $j \in \{ (d_\KK^{\text{weekly}})^{-1} 96 +1, \ldots, (d_\KK^{\text{weekly}})^{-1} 96 +\eta_i^{\text{annual}} -1 \}$ they are iteratively defined 
 as $\wtilde{B}^{\mu^i}_j(t)  = \wtilde{B}^{\mu^i}_{j-1}(t-d_{\KK, i}^{\text{annual}}) $.
Hence the number of basis functions $M(\mu^i) = (d_\KK^{\text{weekly}})^{-1} 96 + \eta^{\text{annual}}_{i} -1$ for $i\in \{\PP, \LL\}$. 

For the wind and solar component we again define the basis functions for the daily and annual cycle as:

$$ \wtilde{B}^{*,\mu^\RR}_{\text{daily},2}(t) = 
  \sum_{k \in  \Z } B_{k S_{\text{daily}}, d_\KK^{\text{daily}}}(t) 
  \text { and }
  \wtilde{B}^{*,\mu^\RR}_{\text{annual},2}(t) = 
   \sum_{k \in   \Z } B_{k S_{\text{annual}} , d_\KK^{\text{annual}}}(t). $$
Please remark that we set  
$\wtilde{B}^{*,\mu^\RR}_{\text{daily},1}(t) = \wtilde{B}^{*,\mu^\RR}_{\text{annual},1}(t) = 1$ which will become necessary in a later step, where we model the interactions.
The subsequent B-splines are again iteratively defined with 
$\wtilde{B}^{*,\mu^\RR}_{\text{daily},h_1}(t) = \wtilde{B}^{*,\mu^\RR}_{\text{daily},h_1}(t - d_\KK^{\text{daily}})$ for $h_1 \in \{ 3, \ldots, \eta^{\text{daily}} \}$ and 
$\wtilde{B}^{*,\mu^\RR}_{\text{annual},h_1}(t) = \wtilde{B}^{*,\mu^\RR}_{\text{annual},h_1}(t - d_\KK^{\text{annual}})$ 
for $h_2 \in \{ 3, \ldots, \eta^{\text{annual}}_\RR \}$. For modeling the interactions of the daily and annual cycle we can now introduce a modified version of basis functions for the renewables:
$$ \wtilde{B}^{*,\mu^\RR}_{j}(t) = 
  \wtilde{B}^{*,\mu^\RR}_{\text{daily},h_1}(t) \wtilde{B}^{*,\mu^\RR}_{\text{annual},h_2}(t),$$
  
where $h_1 =  j \modulo \eta^{\text{daily}} +1 \text{ and } h_2 = j \modulodiv \eta^{\text{annual}}_\RR + 1  $ for $j\in \{1, \ldots, \eta^{\text{daily}} \eta^{\text{annual}}_{\RR} -1\}$. The term mod stands for the modulo operator and div is the integer without the remainder of a division. In this situation our previous definition of the first daily and annual basis functions is used, as the total set of B-spline now contains not only the simple daily and annual basis functions but also the interactions of them. Hence, we have included an amount of $M(\mu^\RR) 
= \eta^{\text{daily}} \eta^{\text{annual}}_{\RR} -1$ basis functions.

Furthermore, in Equation \eqref{eq_main_model}, the parameters $\phi^{i,j}_k(t)$ might also depend on time.
So the autoregressive behavior can also change over time, especially in a seasonal manner. A related approach was also
used by \cite{koopman2007periodic} and \cite{bosco2007deregulated} for modeling daily electricity prices, and by 
\cite{guthrie2007electricity} for intra-day prices.
We assume that this dependence has a structure as in \eqref{eq_mean}, so for $\phi^{i,j}_k(t)$ it is given by
$$ \phi^{i,j}_k(t) = \phi^{i,j}_{k,0} + \sum_{j=1}^{M_i(\phi_k^{i,j})} \phi_{k,j} \wtilde{B}_j^{\phi^{i,j}_k}(t).$$
Again we have to choose the basis functions $\wtilde{B}^{\phi^{i,j}_k}_j(t)$, but from an application point of view it is more practical to 
take the same basis functions as for $\mu^i$, which is implemented in our approach. 

Unfortunately it would expand the parameter space enormously if we assume
that for every possible lag $k\in I_{i,j}$ the coefficient 
$\phi^{i,j}_k(t)$ changes over time. Therefore, we let only
those coefficients $L_{i,j}\subseteq I_{i,j}$ to change over time, which we assume to be most important.
These are summarized in Table \ref{tab_lags_ar},
the other ones are empty.
\begin{table}[tbh]
\centering
\begin{tabular}{r|l}
  \hline
 Index sets & Lags  \\ 
  \hline
$L_{\PP,\PP}$, $L_{\LL,\LL}$ & $1,2,24,25, 168,169$ \\ 
$L_{\RR,\RR}$ & $1,2,23,24,25$ \\ 
\end{tabular}
\caption{Considered lags of the index sets $L_{i,i}$ for $i\in \II$.}
\label{tab_lags_ar}
\end{table}

\begin{figure}[hbt!]
\centering
 \includegraphics[width=0.95\textwidth]{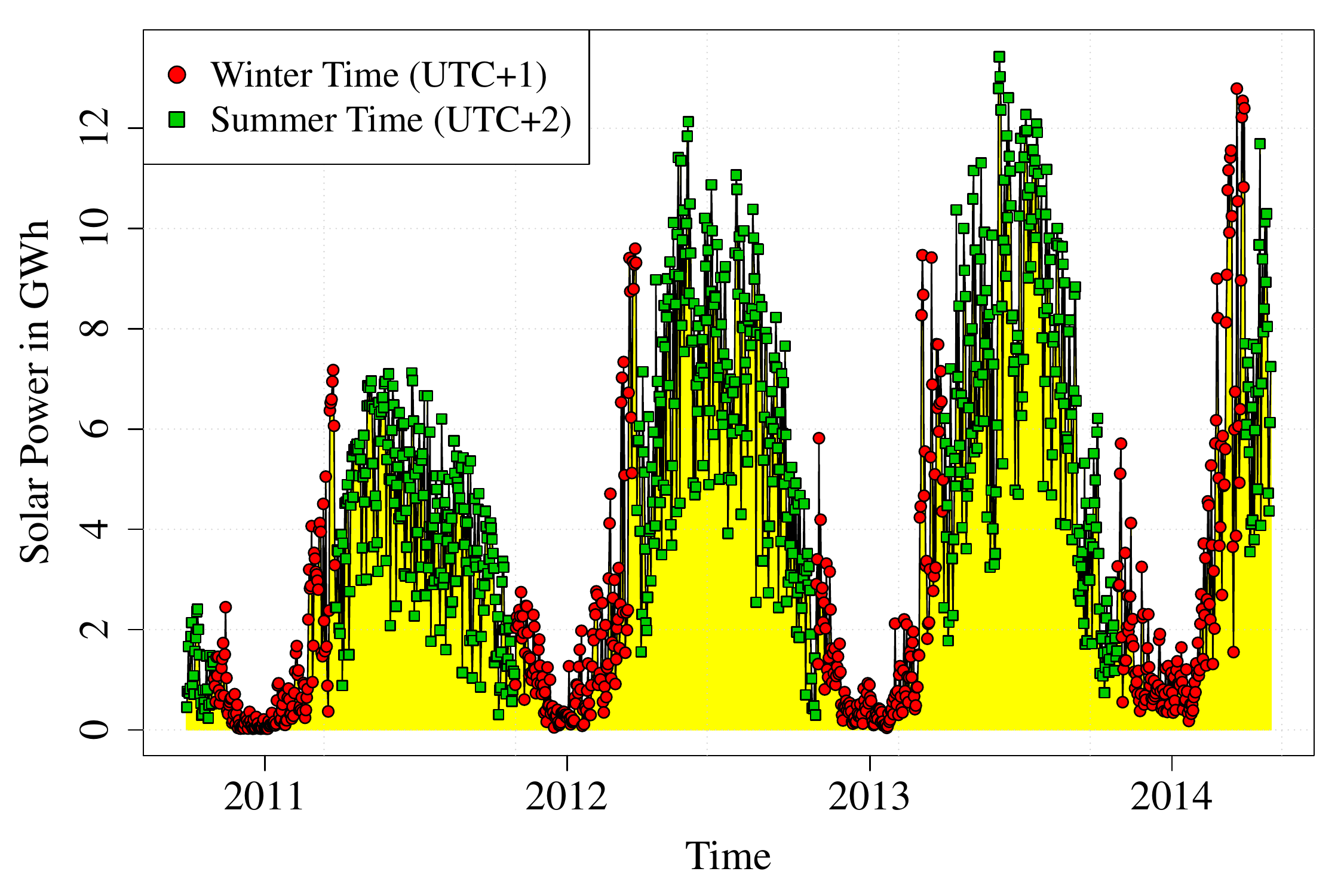}
 \caption{Solar Power feed-in at 8:00 am.}
 \label{fig_solar}
\end{figure}

Furthermore, there is a special impact on $Y_t$ that has not been yet modeled. 
Due to daylight saving time, we have a time shift in March and October. 
As we removed the added hour in October and extrapolated the missing hour in March, there is no problem
in modeling the electricity spot price as the population adjusts their every day behavior very quickly. 
The same holds true for the electricity load if we assume that the influence of regenerative power feed-in is negligible. 
But for the renewable energy feed-in this looks different, here we can improve our model. 
In winter the sun peak is at 12 am, in summer it is at 1 pm. This has an impact
especially on the solar power. Figure \ref{fig_solar} illustrates this effect on the observed solar power feed-in in the morning.
It can be observed that right after the time shift in March there is a 
downwards jump in the solar power feed-in present, and after the time shift in October a jump upwards occurs.
We will model this behavior by a shift in the considered basis functions. 
Therefore, let $\T_{\text{DST}}$ denote the set of all time points with summer time.
Then the considered time adjusted basis functions are given by
$$\wtilde{B}^{\mu^\RR}_{j}(t) = 
\bsone_{\{t\notin \T_{\text{DST}} \}} \wtilde{B}^{*,\mu^\RR}_{j}(t) + \bsone_{\{t\in \T_{\text{DST}} \}} \wtilde{B}^{*,\mu^\RR}_{j}(t+1).$$
This shift is applied to all basis function of the wind and solar power feed-in. 
For the mean component $\mu^{\RR}(t)$ the effect is clear, here this is equivalent to a shift in the hour labels.
In contrast, the effect on the autoregressive parameters is not that obvious. The feed-in 
$Y_{t}^\RR$ depends e.g. on $\phi^{\RR,\RR}_{24}(t)$ before the time shift and is
different to $\phi^{\RR,\RR}_{24}(t)$ after the time shift. 
But $\phi^{\RR,\RR}_{24}(t)$ would be the 
same as $\phi^{\RR,\RR}_{24}(t-25)$ in March or $\phi^{\RR,\RR}_{24}(t-23)$ in October, 
due to the time shift in daylight savings time. Nevertheless, in our approach this is not the case, as we explicitly model the daily and yearly interaction effects. The coefficients therefore truly embed the effects of the time shift.


But also the variance structure of the model is relevant. Here we assume that 
$\Sigma_t = \text{diag}(\bssigma_t) =\text{diag}(\sigma^\PP_t,\sigma^\LL_t,\sigma^\RR_t) $.
The error term $\bseps_t$ can be written as
$$\eps^i_t = \sigma^i_t Z^i_t \text{ where } (Z^i_t)_{t\in \Z} \text{ is i.i.d. with } \E(Z^i_t)=0 \text{ and } \var(Z^i_t) =1 .$$
As modeled by \cite{keles2012comparison} we want to explain conditional heteroscedasticity by an ARCH and GARCH type model.
This approach is suitable to model the price spikes, especially if the underlying distribution of $Z_t$ is heavy tailed.
 
But as the price process 
$\bsY_t$ is heavy tailed due to spikes, the residual process $\bseps_t$ is likely to be heavy tailed with a small tail index, too. 
We noticed that it is not suitable 
to consider the square of $\eps^i_t$, as done in a typical GARCH process, because they become even more heavy tailed. 
This halves the corresponding tail index and the convergence of the estimators will get worse.
Instead of taking squares, we use the absolute value $ |\eps^i_t|$, which keeps the tail index on the same level.
Some GARCH-type processes already use this structure,
e.g. the TARCH model by \cite{rabemananjara1993threshold}. Additionally, a TARCH and TGARCH model can handle the so called leverage effect,
so that negative and positive past residuals have different influences on the volatility.
We assume a TARCH process for $\sigma_t$ so that 
\begin{equation}
\sigma^i_t = \alpha^i_0(t) + \sum_{k \in J_i} \alpha^{+,i}_k \eps^{+,i}_{t-k} + \alpha^{-,i}_k \eps^{-,i}_{t-k} 
\label{eq_arch} 
\end{equation}
holds with  $\alpha^{+,i}_k,\alpha^{-,i}_k\geq 0$ 
where $\eps^{+,i}_t = \max\{\eps^{i}_t, 0\}$ and $\eps^{-,i}_{t} = \max\{-\eps^{i}_t, 0\}$. For the TARCH process we will assume
that only the trend component $\alpha^i_0(t)$ varies seasonally over time. Thus the full 
parameter vector $\bsalpha_i$ is given by $( \alpha^i_{0,0} , \ldots, \alpha^i_{0,M(\alpha^i_{0})} 
, (\alpha^{+,i}_k,\alpha^{-,i}_k)_{k\in J_{i}} )$.
For $\alpha^i_0(t)$ we consider the same basis functions as for
the corresponding $\mu^i(t)$, but without the linear trend, as it could conflict the positivity constraint to the parameters.
Here we make use of another advantage of the periodic B-spline approach in contrast to a Fourier approximation.
The positivity of the considered basis functions gives us automatically the restriction that the corresponding parameters
have to be positive as well as $\sigma_t$ must be positive.

The index sets $J_i$ that we use are given in Table \ref{tab_lags_eps}, they contain the typical discussed lags.

\begin{table}[tbh]
\centering
\begin{tabular}{r|c}
  \hline
 Index sets & Lags  \\ 
  \hline
$J_{\PP}$, $J_{\LL}$ & 1:361, 504:505, 672:673, 840:841\\ 
$J_{\RR}$ & 1:361 \\ 
\end{tabular}
\caption{Considered lags of the index sets $J_i$ with $i\in \II$ for the TARCH part.}
\label{tab_lags_eps}
\end{table}

By multiplying $\gamma^i = \E|Z^i_t|$  and adding
$v^i_t = \sigma^i_t(|Z^i_t| - \gamma^i)$ on both sides of equation \eqref{eq_arch} we get 
\begin{equation}
|\eps^i_t| = \tilde{\alpha}^i_0(t) +  \sum_{k \in J_i} \tilde{\alpha}^{+,i}_k \eps^{+,i}_{t-k} 
+ \tilde{\alpha}^{-,i}_k \eps^{-,i}_{t-k} + v^i_t,
\label{eq_absarch} 
\end{equation}
where $\tilde{\alpha}^{i}_0(t) = \gamma^i \alpha^{i}_0(t)$, 
$\tilde{\alpha}^{+,i}_k = \gamma^i \alpha^{+,i}_k$, and $\tilde{\alpha}^{-,i}_k = \gamma^i \alpha^{-,i}_k$.
Here $v^i_t$ is a weak white noise process with $\E(v^i_t) = 0$.
The auxiliary model in \eqref{eq_absarch} is useful as it allows us to estimate $\bsalpha_i$ given
estimates of $\bseps^i$.
The fitted values $\tilde{\sigma}^i_t$ 
of equation \eqref{eq_absarch} are proportional to $\sigma^i_t$ up to the constant $\gamma^i$.
  $\gamma^i$ is the first absolute moment 
 of $\eps^i_t$ which might help to characterize the distribution of $\eps^i_t$. If $\eps^i_t$ follows a normal distribution it is exactly
 $\sqrt{2\pi^{-1}} \approx 0.798$, whereas, e.g., the standardized t-distributions have a larger first absolute moment.

Given reasonable estimates $\what{\eps}^i_t$ and $\what{\tilde{\sigma}}^i_t$ for 
$\eps^i_t$ and $\tilde{\sigma}^i_t$ we can use the plug-in estimator for $\gamma^i$ to estimate it.
With $\kappa^i_t = (\what{\tilde{\sigma}}^i_t)^{-1} \what{\eps}^i_t$ this is given by 
 \begin{equation}
  \what{\gamma}^i =   \left(\frac{1}{n-1}\sum_{t=1}^n (\kappa^i_t - \overline{\kappa^i})^2 \right)^{-\frac{1}{2}} ,
  \label{eq_est_gamma}
 \end{equation} 
  as $\var(\kappa_t^i)=({\gamma^i})^{-2}$.
 $\sigma^i_t$ can now be estimated with $\what{\sigma}^i_t = (\what{\gamma}^i)^{-1} \what{\tilde{\sigma}}^i_t $.

Finally, we can rewrite \eqref{eq_main_model} in matrix notation with
\begin{equation}
\bsY^i = \bsX^i \bstheta_i + \bseps^i 
\label{eq_main_model_matrixform}
\end{equation}
with $\bsY^i = (Y_1^i, \ldots, Y^i_n)$, $\bsX^i = (\bsX^{i}_1, \ldots, \bsX^i_{n})$,  
$\bseps^i = (\eps^i_1, \ldots, \eps^i_n)$ and parameter vector $\bstheta_i$
for $i\in \II$ given time points $t\in \{1,\ldots, n\}$.
The vectors $\bsX_t^i$ are given through equation (\ref{eq_mean}) by
$$\bsX_t^i = \left( 1, t, \bsB_t(\mu^i), \bsB^{Y^i}_t(I_{i,\PP}) , \bsB^{Y^i}_t(I_{i,\LL}) ,\bsB^{Y^i}_t(I_{i,\RR}) ,
\bsB^{Y^i, \text{per}}_{t}(L_{i,i}) \right)$$
where we have $\bsB_t(\mu^i) = (\wtilde{B}^{\mu^i}_1(t), \ldots, \wtilde{B}^{\mu^i}_{M(\mu^i)}(t) )$,
$\bsB^{Y^i}_t(I_{i,j}) = (Y^i_{t-k})_{k\in I_{i,j}}$, for $j \in \II$, and 
$\bsB^{Y^i,\text{per}}_t(L_{i,i}) = (Y^i_{t-k} \bsB_t(\mu^i) )_{k\in L_{i,i}} $.
In the following we will refer the parameters that correspond to $(1, t, \bsB_t(\mu^i))$ 
as deterministic component, $\bsB^{Y^i}_t(I_{i,\PP})$ as price component, 
$\bsB^{Y^i}_t(I_{i,\LL})$ as load component, 
$\bsB^{Y^i}_t(I_{i,\RR})$ as wind and solar component and $\bsB^{Y^i, \text{per}}_{t}(L_{i,i})$
 as periodic component
given an $i\in \II$. 

All in all our model has approximately 3500 possible parameters, 
which are given in Table \ref{tab_summary_parameters}. 
They match the mentioned deterministic, price, load, wind and solar, and periodic component defined above. Please notice that these parameters only represent the maximum amount of possible parameters, as our estimation procedure, which will be explained in the next section, will eliminate parameters, which are not significantly different from zero.

\begin{table}[tbh]
\centering
\begin{tabular}{p{21mm}|llllp{23mm}|l||c}
  \hline
 number of parameters & $\mu^i(t)$ & $I_{i,\PP}$ & $I_{i,\LL}$ & $I_{i,\RR}$ & seasonal $\phi^{i,j}$ coefficients & sum ($p_i$) & $\bsalpha_i$ (TARCH)  \\   \hline
$ i = \PP$ & 
36 & 369 & 369 & 49 & $6\times35$ & 1033 & 402  \\ 
$ i = \LL$ & 
36 & - & 369 & 49 & $6\times35$ & 664 & 402 \\ 
$ i = \RR$ & 
37 & - & - & 361 & $5\times36$ & 578 & 397 \\ 
\end{tabular}
\caption{Summary table of the amount of used parameters $p_i$ of the model 
given in \eqref{eq_main_model} and the TARCH model in equation \eqref{eq_arch}  $i\in \II$.}
\label{tab_summary_parameters}
\end{table}

\section{Estimation method} \label{Estimation}

For the estimation of the parameters we will consider a variation of the iteratively weighted least squares approach as used for example 
in \cite{mbamalu1993load} or \cite{mak1997estimation}. So basically we apply a weighted least squares estimation of
 model \eqref{eq_main_model}. Then we use 
the corresponding residuals to estimate the volatility which is used to compute new weights for
re-estimating model \eqref{eq_main_model}.

The considered model has a lot of regressors, 
and it is likely that some of them have no significant impact on the model, but rather increase 
spurious effects. 
The suggested lasso 
(least absolute shrinkage and selection operator) approach can handle this in an efficient way.
It was introduced by \cite{tibshirani1996regression} in the context of shrinkage and selection in regression models
and was recently applied by \cite{hsu2008subset} and \cite{ren2010subset} in the context 
of vector autoregressive models. 

The weighted lasso estimator of the parameter vector $\bstheta^i$ of equation \eqref{eq_main_model} or \eqref{eq_main_model_matrixform}
is given by
\begin{equation}
 \what{\bstheta_i} = \argmin_{\theta \in \R^{p_i}} 
 \sum_{t=1}^{n} (Y^i_t - w^i_t \bsX_t^i \bstheta^i)^2 + \lambda_{i,n} \sum_{j=1}^{p_i} | \theta_j |
\end{equation}
with a weight vector $\bsw^i = (w^i_1, \ldots, w^i_n)$, $p_i$ as length of $\bstheta^i$ 
and tuning parameter $\lambda_{i,n}$.
In the initial step $\bsw^i$ is chosen to be $\bsone$. 
For the estimation we use the least angle regression (LARS) estimation algorithm of \cite{efron2004least}.

As selection criterion we consider the Akaike information criterion (AIC) to estimate . This is asymptotically 
equivalent to the cross-validation in regression analysis as suggested in \cite{efron2004least}

In the next step we want to compute a new weight matrix $\bsW = (\bsw^1, \ldots, \bsw^d)$
based on the volatility estimates for $\bssigma^i_t$.
Thus we have to estimate the volatility parameters
from equation \eqref{eq_arch} using equation \eqref{eq_absarch} and \eqref{eq_est_gamma}.

Here we have to ensure that all parameters take non-negative values, as the regressors are all positive.
To solve this non-negative least squares problem, we use the
NNLS algorithm as described by \cite{lawson1974solving}. Note that this estimation technique can 
be seen as a parameter selection approach as well,
as some parameters are likely estimated to be 0. 
This sparsity effect in a NNLS problem was recently analyzed by \cite{meinshausen2013sign}.
\cite{slawski2013non} showed that the non-negative least squares approach is potentially 
superior to the positive lasso, a lasso approach with a positive parameter constraint. 

The general estimation scheme is given by

\framebox{
\parbox{.89\textwidth}{
\begin{centering}
\begin{enumerate}
 \item[\textit{1)}] Set the initial $d\times n$ dimensional weight matrix $\bsW = (\bsone, \ldots, \bsone)$ and 
 the iteration parameter $K=1$.
 \item[\textit{2)}] Estimate \eqref{eq_main_model} using LARS-lasso method with weights $\bsW$.
 \item[\textit{3)}] Estimate $\bssigma_t$ by \eqref{eq_absarch} and \eqref{eq_est_gamma} with 
 $|\what{\bseps}_t|$ as absolute residuals from \textit{2)}\\
 using the NNLS algorithm.
 \item[\textit{4)}] Redefine $\bsW = (\bsw^1, \ldots, \bsw^d)$ by $ \bsw^i= \left( {\left( \what{\sigma}^i_1\right)}^{-2}, \ldots ,
 {\left( \what{\sigma}^i_n\right)}^{-2} \right)$ \\ with $\what{\bssigma}_t = 
 \left(\what{\sigma}^1_t,\ldots,\what{\sigma}^d_t\right)$ as fitted values from \textit{3)}.
 \item[\textit{5)}] Stop the algorithm, if a stopping criteria is satisfied, \\ otherwise $K=K+1$ and go back to \textit{2)}.
 \end{enumerate}
\end{centering}
\vspace{-5mm}
}
}
\vspace{2mm}

Note that before performing the LARS algorithm 
on the weighted regressor matrix, we standardize their columns so that the columns of $\bsX^i$ from equation 
\eqref{eq_main_model_matrixform} have zero sample mean
and sample variance $1$, so that LASSO can act as a proper model selection algorithm.
Additionally we could also replace $\what{\bssigma}_t$ by $\what{\tilde{\bssigma}}_t$ as they 
are the same upto the constant $\what{\gamma}_i$. This does not change the result of the estimation algorithm,
but we can improve the computation time slighty as we do not have to estimate $\gamma^{i}$ by equation \eqref{eq_est_gamma}.

\begin{figure}[hbt!]
\centering
 \includegraphics[width=0.69\textwidth]{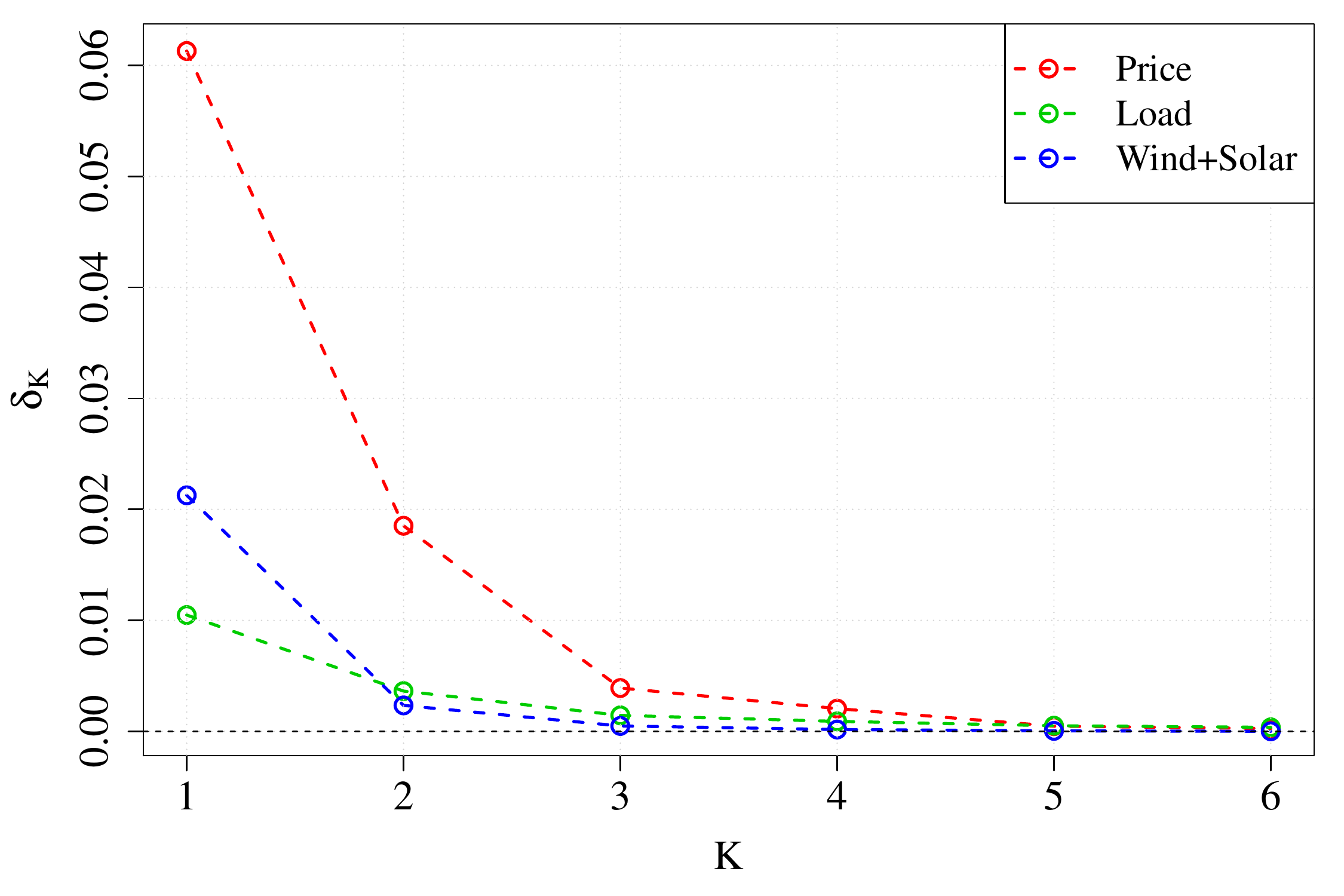}
 \caption{$\delta_K^i$ for iteration $K\in \{1,\ldots,6\}$ and $i\in \II$, estimated on the full available sample.}
 \label{fig_sigmaconv}
\end{figure}

For the stopping criteria we suggest to look at the convergence of the $\bssigma^i$ within the algorithm.
A reasonable criteria is to iterate until $\delta_K^i = \| \what{\bssigma}^i_{K-1} - \what{\bssigma}^i_{K}\| <\epsilon$ for all $i\in \II$.
For 
$K=1$ the $\bssigma^i_{1}$ is the homoscedastic estimate from the homoscedastic lasso estimation. 
For our application 
we choose $\delta_K^i$ as $n^{-1} \|\what{\bssigma}^i_{K-1} - \what{\bssigma}^i_{K}\|_1$ 
and $\epsilon=0.001$.  
In the empirical study we noticed that this algorithm converges quickly (see Figure \ref{fig_sigmaconv}), so 
a small number of iterations $K_{\text{max}}$ seems to be sufficient for practitioners.
In Figure \ref{fig_sigmaconv} we can observe that in the fifth iteration the change in $\what{\bssigma}^i$ is 
less than $\epsilon$. Hence, we decide to stop after $K_{\text{max}} = 4$ iterations in the rest of our study. 

The estimation technique used has the huge advantage of being fast, even if hundreds of potential regressors are included in the model. 
The computational complexity of the dominating LARS algorithm is the same as for a common OLS estimation, which is 
$\OO(K_{\text{max}} dp^2n)$ for 
$p<n$ where $K_{\text{max}}$ is the number of maximal iteration steps,
$d$ the dimension of the process, $p$ the maximal number of regressors in
\eqref{eq_main_model_matrixform}, and $n$ the number of observations. Here we see that the number of parameters $p$ in 
the models has a quadratic impact, thus it is important to keep the parameter space manageable regarding the computation time. 
In our computations the iteratively reweighted lasso with $K_{\text{max}}=4$ takes only a few minutes on a 3 GHz computer,
with $n$ about 31,000, $p$ about 3500, and $d=3$.

The consistency and asymptotic behavior of $\what{\bstheta}_i$ and $\what{\bsalpha}_i$ for the process considered 
is not obvious.
However, the lasso-type estimators in a heteroscedastic regression setting  
 were recently analyzed in \cite{severien2012shrinkage}, \cite{wagener2012bridge}, and \cite{wagener2013adaptive}.

There are also some asymptotic results for the lasso estimator in an autoregressive setting.
Important results are given for univariate ARX processes by \cite{wang2007regression}
and for the VAR model by \cite{hsu2008subset}. 

Even though there is a lack of theoretical justification for the asymptotic properties of the used estimators, we will
assume asymptotic normality for $\what{\theta}_i$ and $\what{\bsalpha}_i$.
For $\what{\theta}_i$ we assume the asymptotic behavior as in \cite{wagener2012bridge}. 
Following their result 
we have $\sqrt{n} (\what{\bstheta}_i(1) -\bstheta_i^*(1)) \to N(\bszeta_i(\bstheta_i^*), \bsGamma_i(\bstheta_i^*) )$ in distribution 
where $\bstheta_i^*$ is the true parameter vector, $\what{\bstheta_i}(1)$ and $\bstheta_i^*(1)$ the corresponding non-zero parts,
$\bszeta_i(\bstheta_i^*)$ as asymptotic mean vector,
and $\bsGamma_i(\bstheta_i^*)$ the asymptotic covariance matrix, if $n^{-\frac{1}{2}} \lambda_{i,n} \to \lambda_{i,0} \geq 0$.
For the mean and the covariance matrix we have 
 $\bszeta_i(\bstheta_i^*) = - \frac{\lambda_{i,0}}{2} \bsG_i^{-1} \text{sign}( \bstheta_i^*(1) ) $
and $\bsGamma_i(\bstheta_i^*) = \bsG_i^{-1} \bsG_i^{\bsW} \bsG_i^{-1}$
 with $\bsG_i$ as limit of
the Gramian $\bsG_{n,i} = n^{-1}\bsX_i^\top \bsX_i$, $\bsG_i^{\bsW}$
as limit of the weighted Gramian $\bsG^{\bsW}_{n,i} = n^{-1}\bsX_i^\top \bsW_i \bsX_i$, 
weight matrix $\bsW_i = \text{diag}({(\sigma_1^i)}^{-2},\ldots, {(\sigma_n^i)}^{-2})$, and $\text{sign}$ as signum function.

\begin{figure}[hbt!]
\centering
 \includegraphics[width=0.69\textwidth]{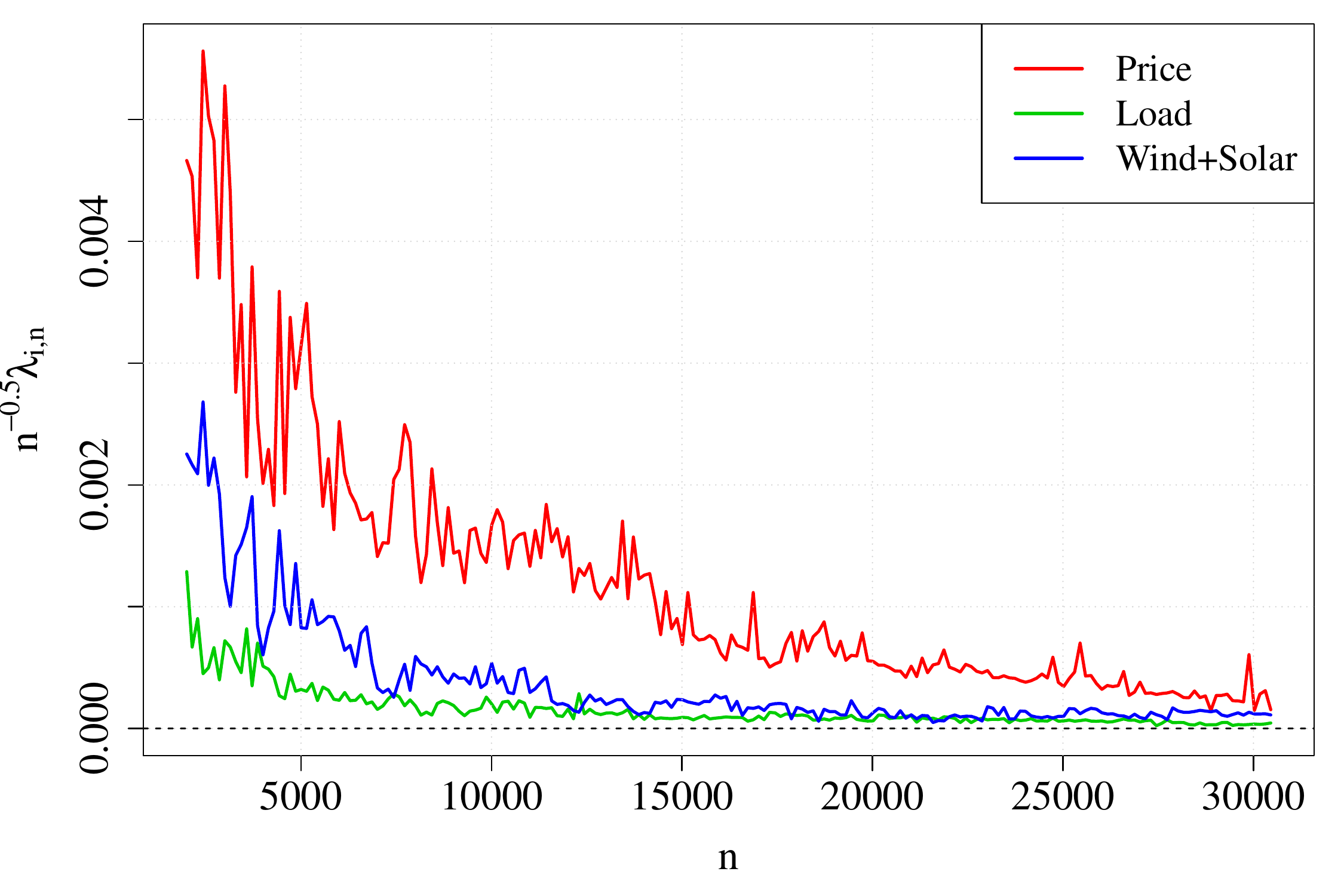}
 \caption{$n^{-\frac{1}{2}}\lambda_{i,n}$ for a selection of various $n$.}
 \label{fig_lambdan}
\end{figure}

We see that the asymptotic behavior depends linear on the unknown limit $\lambda_{i,0}$.
For analyzing the asymptotic in our situation we estimated the model for many different $n$ 
ranging from $2000$ to our maximum sample size
and computed the corresponding $\lambda_{n,i}$. In Figure \ref{fig_lambdan} we plotted
$n$ against $n^{-\frac{1}{2}}\lambda_{i,n}$.
There we can observe that $n^{-\frac{1}{2}}\lambda_{i,n}$ goes clearly to zero as $n$ increases. Thus
the assumption $\lambda_0=0$ seems to be appropriate, and the zero vector $\bsnull$ is a reasonable choice
as an estimator for the mean vector $\bszeta_i(\bstheta_i^*)$.
In addition the covariance matrix can be estimated by the common plug-in estimator
$ \what{\bsGamma}_i = 
 n(  \bsX_i^\top \bsX_i  )^{-1} \bsX_i^\top \what{\bsW}_i \bsX_i  ( \bsX_i^\top \bsX_i )^{-1}$.

\section{Results} \label{Results}
\begin{figure}[hbt!]
\centering
\begin{subfigure}[b]{.49\textwidth}
 \includegraphics[width=1\textwidth]{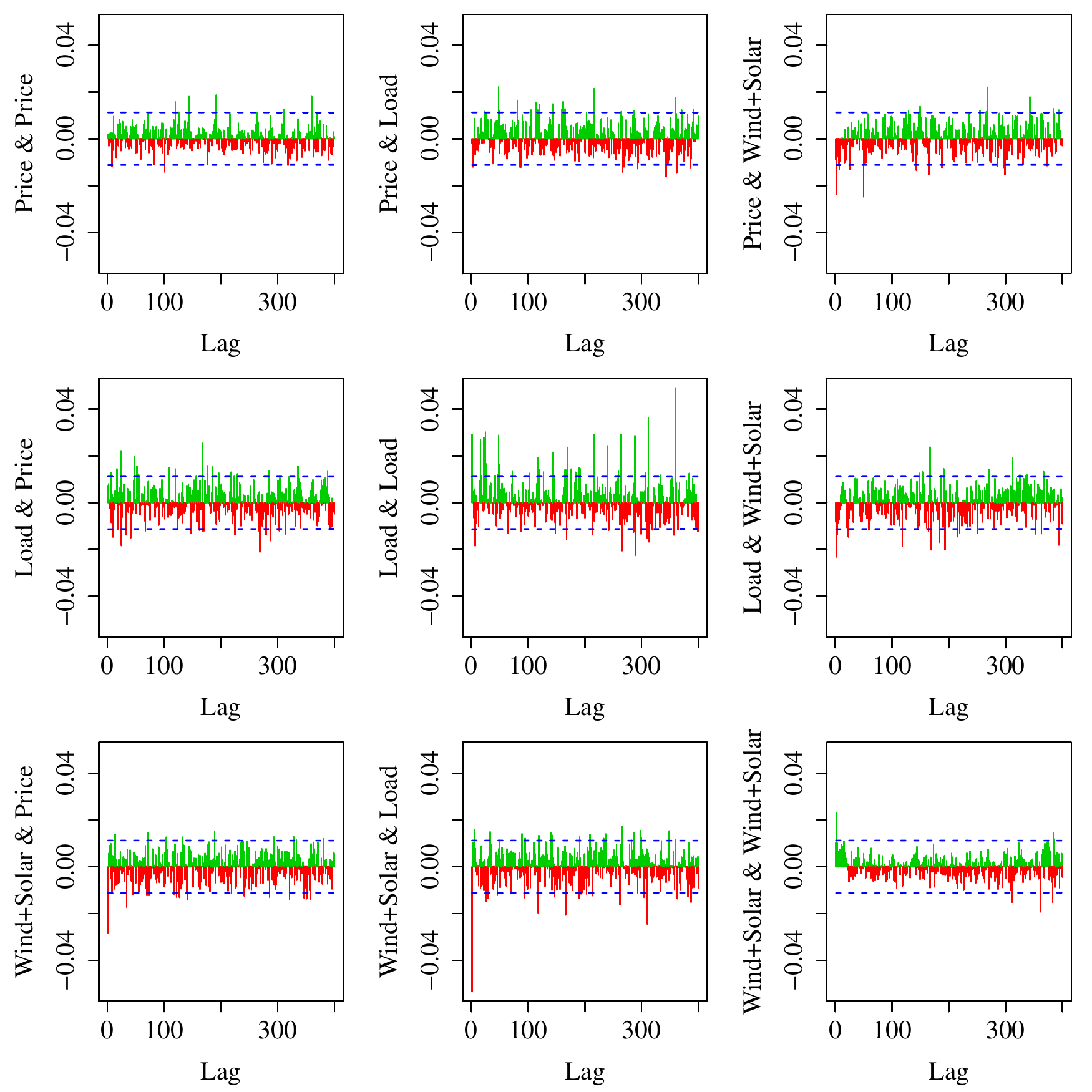} 
  \caption{ACF of $\what{\bsZ}_t$}
  \label{fig_acf_sub1}
\end{subfigure}
\begin{subfigure}[b]{.49\textwidth}
 \includegraphics[width=1\textwidth]{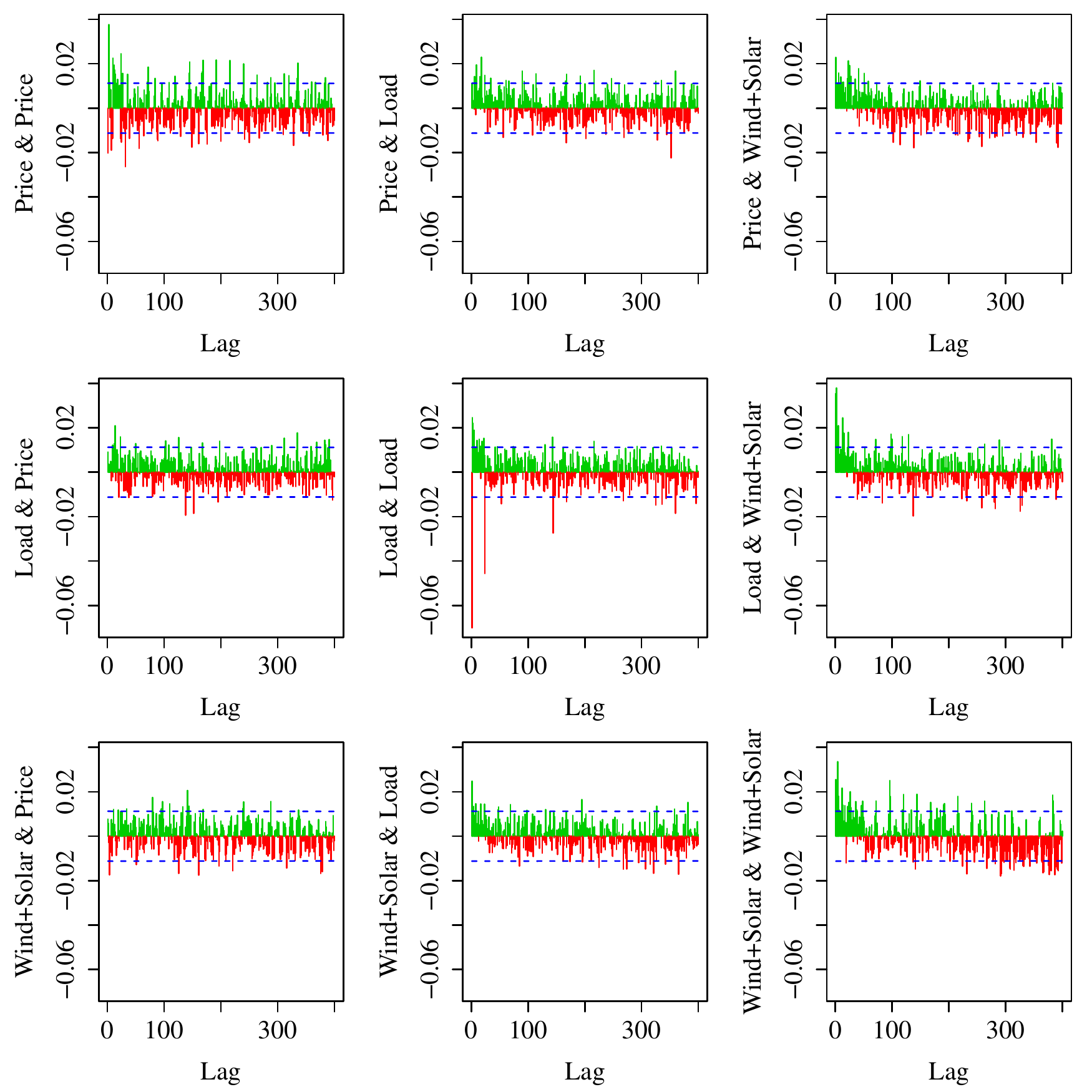} 
  \caption{ACF of $|\what{\bsZ}_t|$}
  \label{fig_acf_sub2}
\end{subfigure}
 \caption{Sample autocorrelation function of $\what{\bsZ}_t$ (\ref{fig_acf_sub1}) and $|\what{\bsZ}_t|$ (\ref{fig_acf_sub2}). }
 \label{fig_acf_res}
\end{figure}

We performed the mentioned estimation technique on the full data set.
For the model diagnostic the sample autocorrelation function (ACF) of the estimated standardized 
residuals and their absolute values are given
in Figure \ref{fig_acf_res}. 

The left 3x3 matrix depicts how the standardized residuals of one time series correlate with the 
lagged standardized residuals of the same or any other of the three time series. The right 3x3 matrix shows
the same relation for the absolute standardized residuals. For instance, the left upper ACF shows the correlation
of the standardized residuals of the price with its own lagged standardized residuals. The illustration right next to it,
Price \& Load, shows the correlation of the standardized residuals of the price with the lagged standardized residuals of the load.
Hence, we can see that the independence assumption for $\bsZ_t$ seems to be satisfied for most of the relations. 
However, the strongest serial correlation structure can be observed for the standardized residuals $Z_t^{\LL}$ of the load.
There could be several reasons for the remaining correlation structure.
For example, we may simply did not choose enough parameters to cover the complex dependence structure,
so an enlargement of $I_{\LL,\LL}$ or $L_{\LL,\LL}$ or a reduction of $d_\KK$ might help.
It could be also possible that there might be non-linear or interactions effects that were not considered so far. 

Further, the sample autocorrelation of $|\bsZ_t|$ is mainly zero, so that $\bsZ_t$ seems to be quite homoscedastic. Thus
the TARCH approach seems to be appropriate to explain the conditional heteroscedasticity. 
Nevertheless, there is still space for improvements.

\begin{figure}[hbt!]
\centering
\begin{subfigure}[b]{0.325\textwidth}
 \includegraphics[width=1\textwidth]{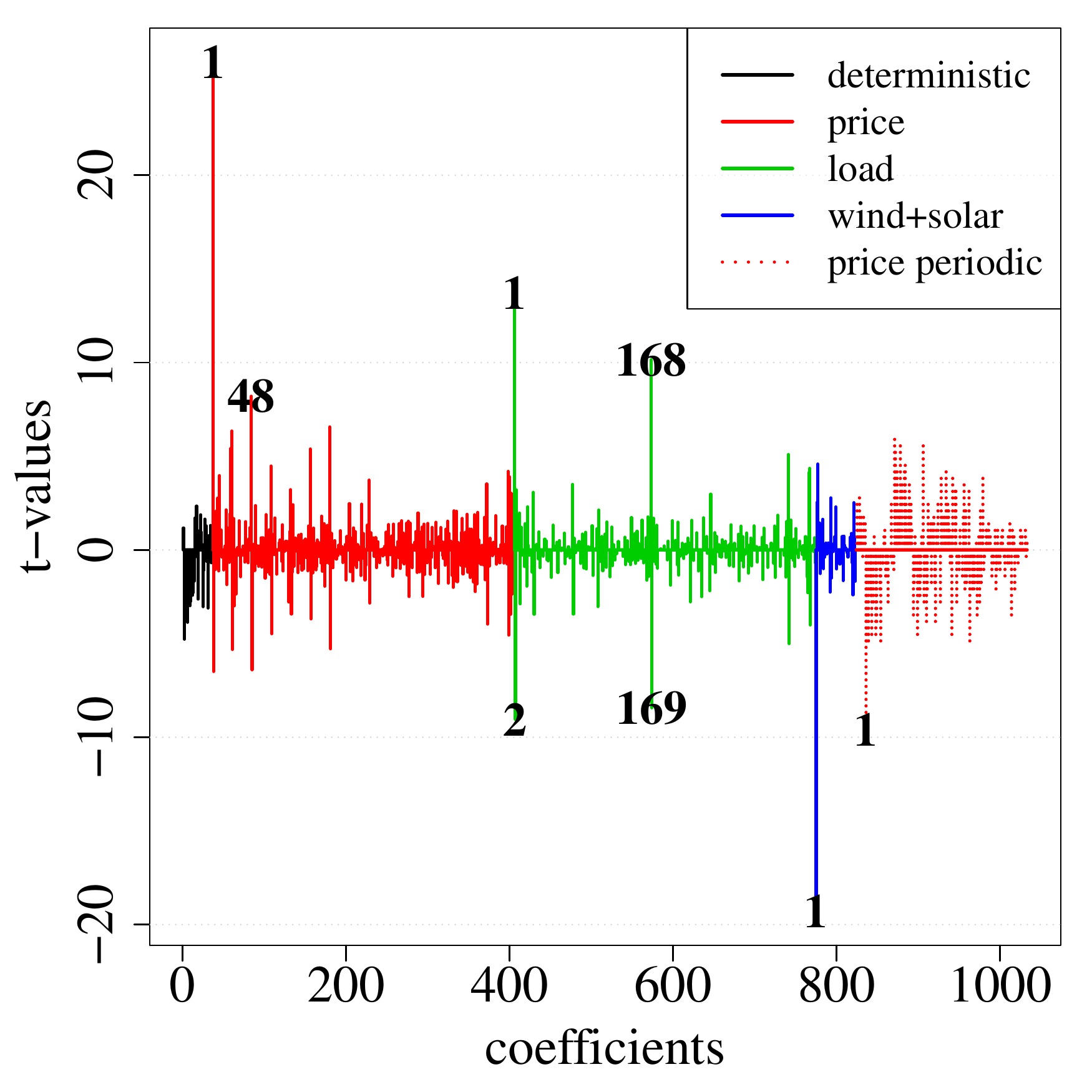}
  \caption{t-values for $Y_t^\PP$}
  \label{fig_tval_sub1}
\end{subfigure}
\begin{subfigure}[b]{0.325\textwidth}
 \includegraphics[width=1\textwidth]{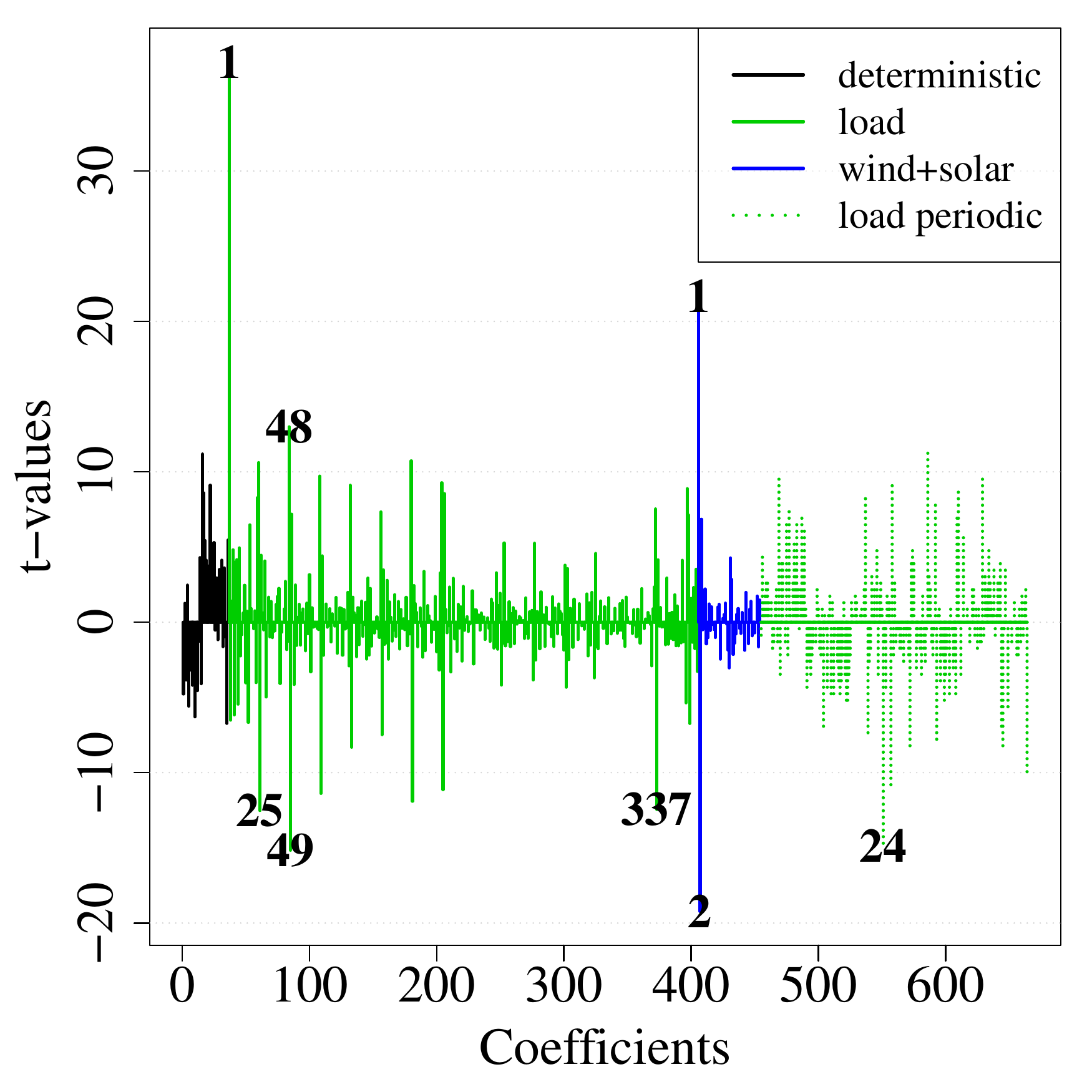}
  \caption{t-values for $Y_t^\LL$}
  \label{fig_tval_sub2}
\end{subfigure}
\begin{subfigure}[b]{0.325\textwidth}
 \includegraphics[width=1\textwidth]{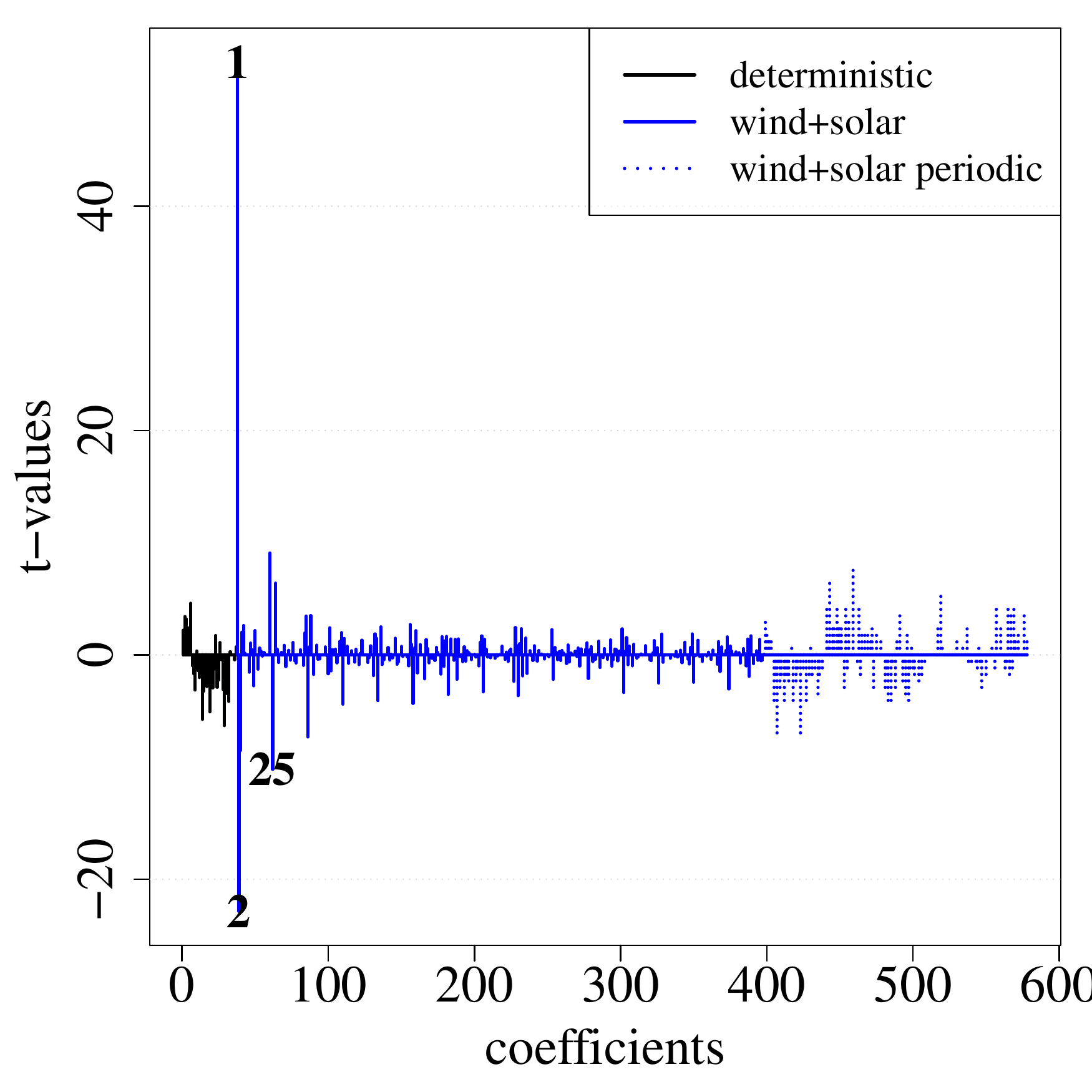}
  \caption{t-values for $Y_t^\RR$}
  \label{fig_tval_sub3}
\end{subfigure}
 \caption{t-values of the estimated model for $\bsY_t$ distinguished by deterministic, price, load, wind+solar and periodic coefficients.}
 \label{fig_tvals}
\end{figure}

Given the estimated model \eqref{eq_main_model} we can evaluate the t-values of every coefficient 
given by their estimate over its standard error which can be estimated under the asymptotic normality assumption.
Roughly speaking we can say, that the larger the absolute t-value
of a coefficient, the more significant is its impact on the process. 
The evaluated t-values are given in Figure \ref{fig_tvals}, distinguished into the groups
introduced at the end of section \ref{Modeling electricity prices}.  

It is worth mentioning that the load and wind and solar feed-in seem to have a strong impact on
the actual electricity price, $Y_t^\PP$.
The most recent past load values (lag 1 and 2) seem to be very important as well as the values one week ago
(lag 168, 169). In contrast to this, wind and solar power has a vast impact only in the short-run at lag 1. 
For the load dependency structure, $Y_t^\LL$, we receive that the first autoregressive lag and the subsequent daily lags tend to be of great importance. Interestingly, lag 337, which accounts for effects two weeks ago, seem to also have a special importance. 
But there is also a strong relationship concerning the renewable energy feed-in, which is especially high for lag 1 and 2.
For the solar and wind power feed-in, $Y_t^\RR$, it is interesting that the first two autoregressive lags clearly dominate all the other larger lags. However, the daily lagged variables still play an important role.

In general we can observe that the dependency structure is quite complex. Therefore we conclude that the chosen set of lags $I_{i,j}$ and $L_{i,i}$, though having thousands of parameters, seem to be reasonable. This is also backed by the results of the parameter selection algorithm.
For the price process $82.2\%$ of the chosen parameters were included. The load process was set to $92.8\%$ and wind and solar to $76.6\%$ of the maximum amount of parameters. Especially the high percentage for the load indicates that an enlargement of the parameter set
is likely to improve the results.

\begin{figure}[hbt!]
\centering
 \includegraphics[width=0.49\textwidth]{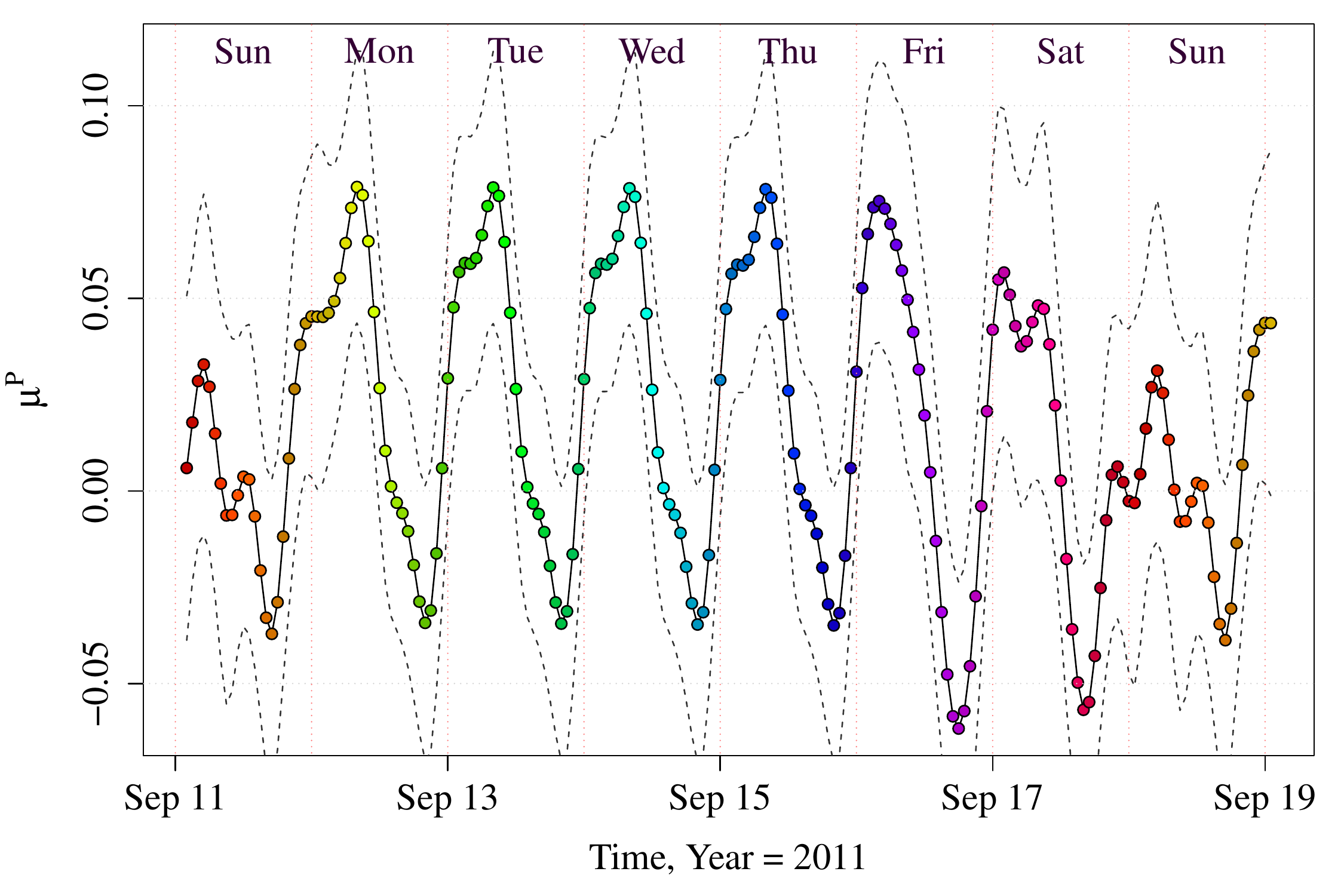}
 \includegraphics[width=0.49\textwidth]{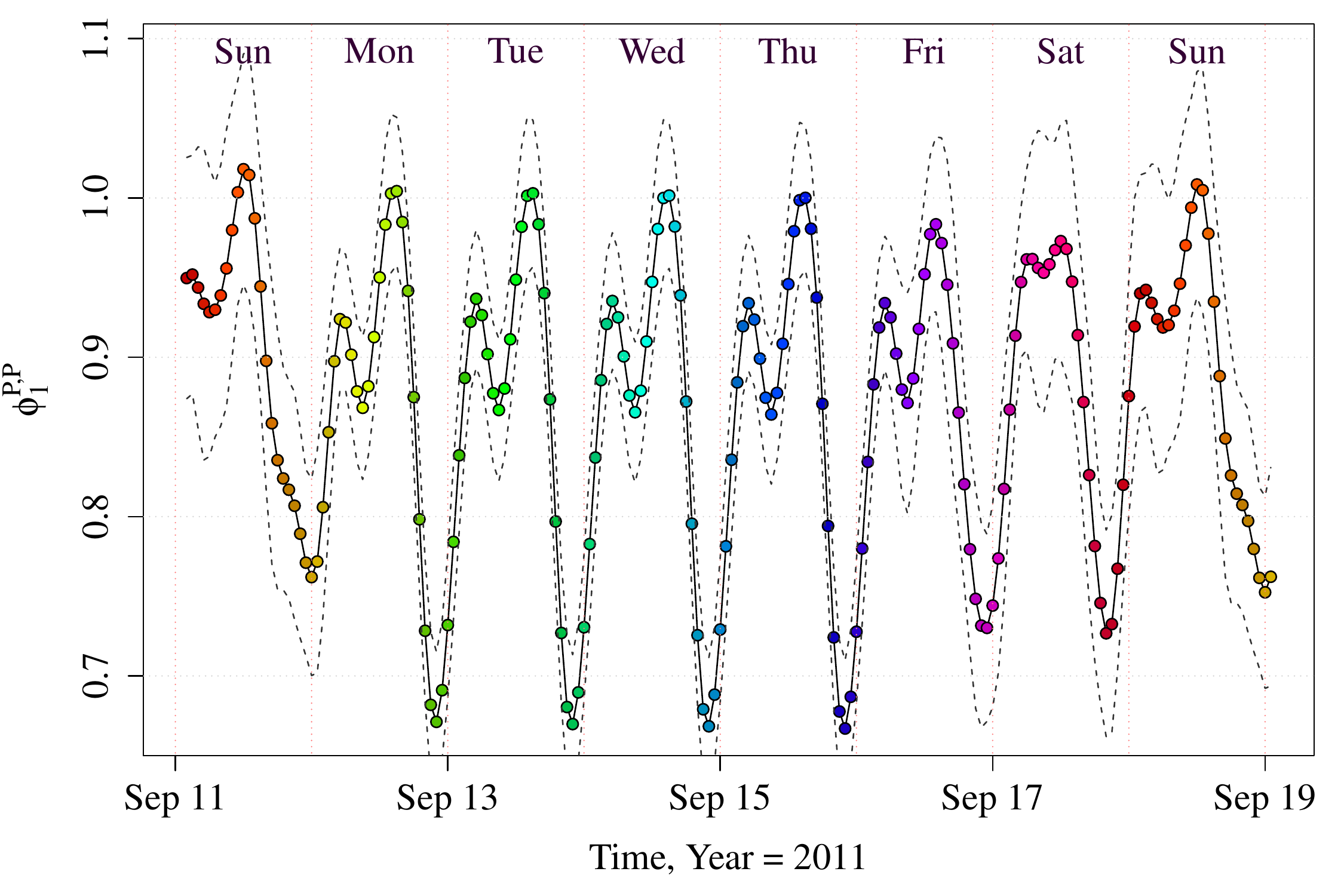}
 \caption{$\what{\mu}^\PP(t)$ and $\what{\phi}^{\PP,\PP}_1(t)$
 within 8 days with approximate $95\%$ confidence intervals.}
 \label{fig_results_periodic1}
\end{figure}

Assuming asymptotic normality allows us to compute several confidence intervals and bands. For example, 
for $\bsmu(t)$ and time varying autoregressive parameters $\phi^{i,j}_k(t)$. We illustrate
the behavior of $\mu^\PP(t)$ and $\phi^{\PP,\PP}_1(t)$
in Figure \ref{fig_results_periodic1}.
From this illustration it can be seen that all coefficients have distinct weekly patterns and a seasonal structure.

\begin{figure}[hbt!]
\centering
\begin{subfigure}[b]{0.325\textwidth}
 \includegraphics[width=1\textwidth]{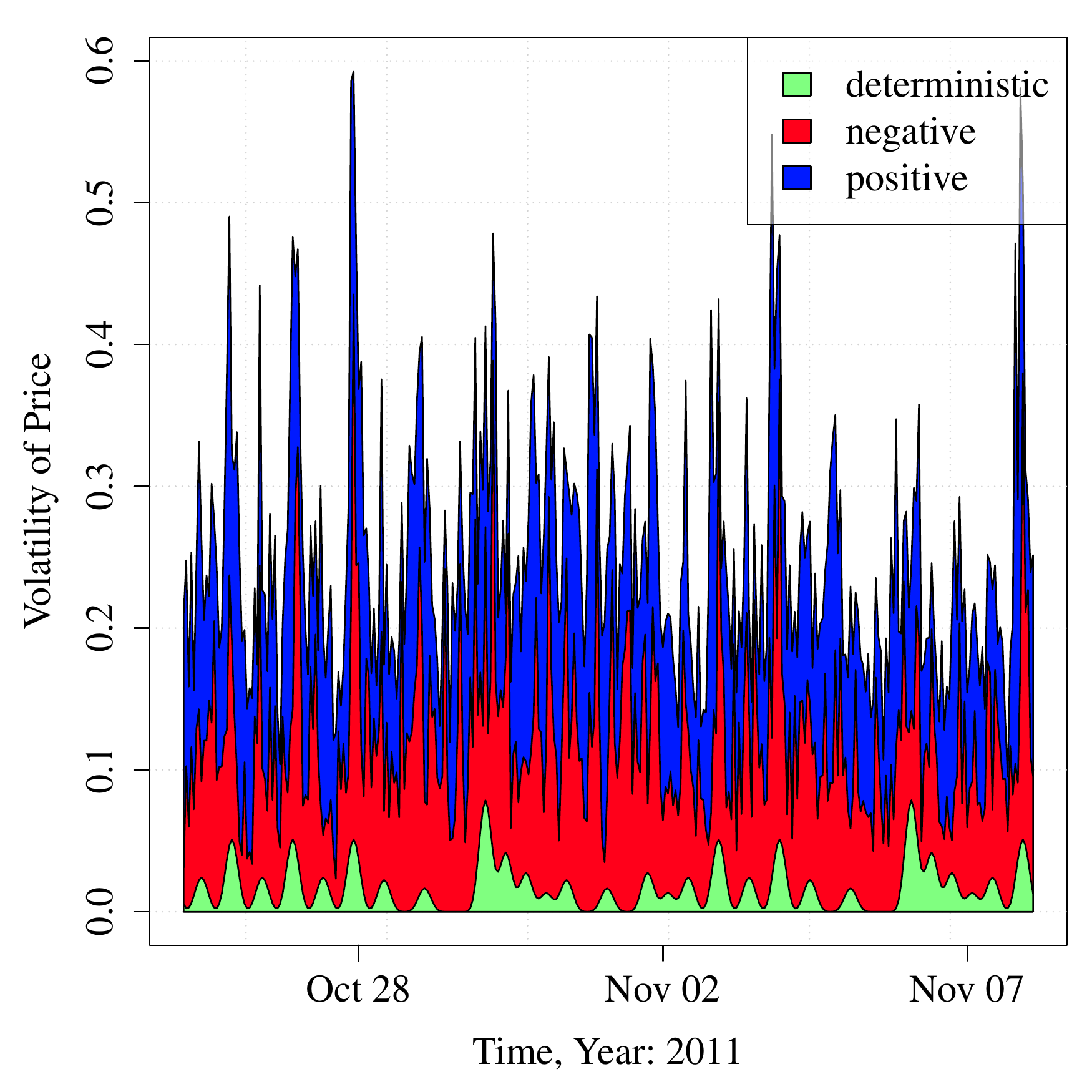}
  \caption{Decomposition of $\what{\sigma}_t^\PP$}
  \label{fig_volatility_sub1}
\end{subfigure}
\begin{subfigure}[b]{0.325\textwidth}
 \includegraphics[width=1\textwidth]{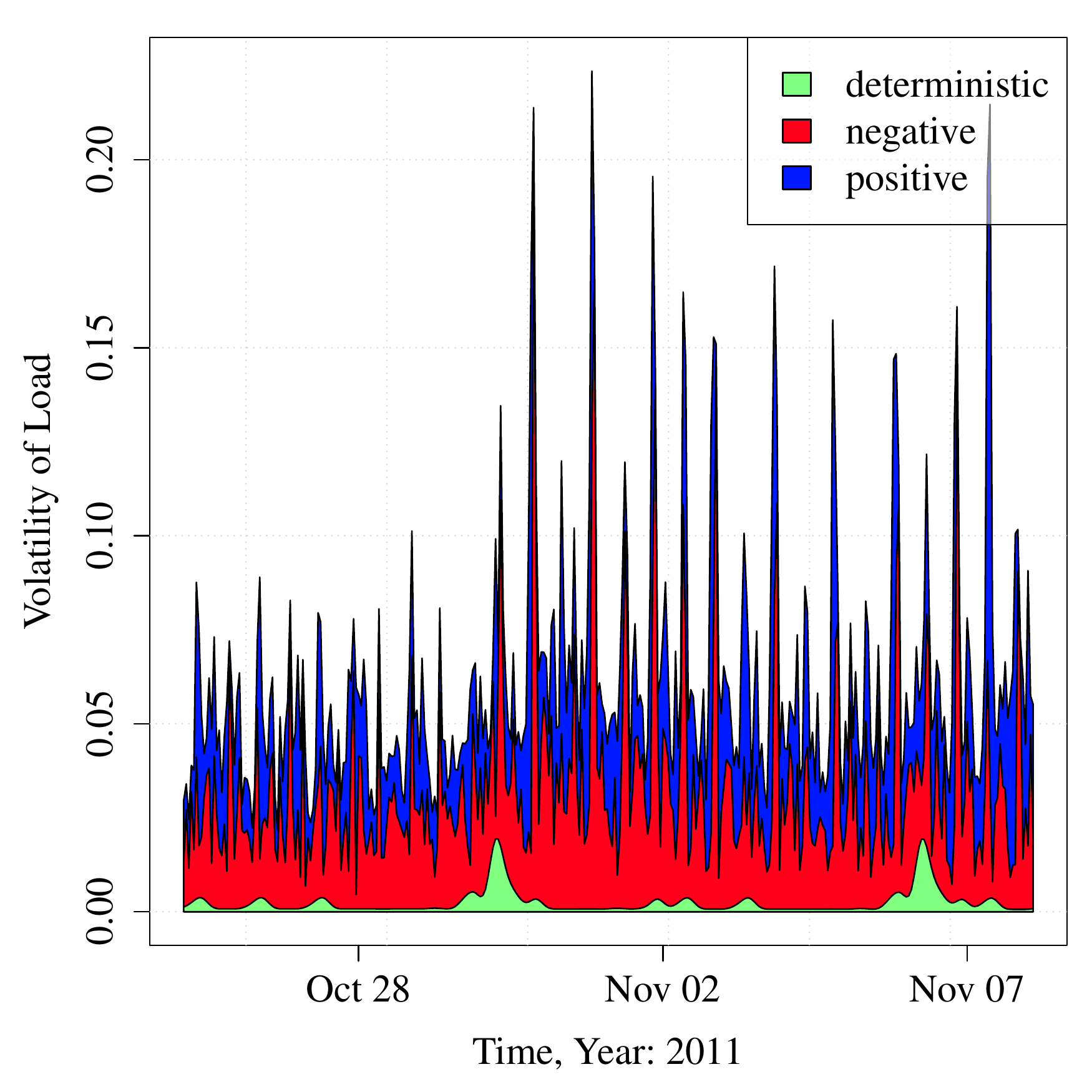}
  \caption{Decomposition of $\what{\sigma}_t^\LL$}
  \label{fig_volatility_sub2}
\end{subfigure}
\begin{subfigure}[b]{0.325\textwidth}
 \includegraphics[width=1\textwidth]{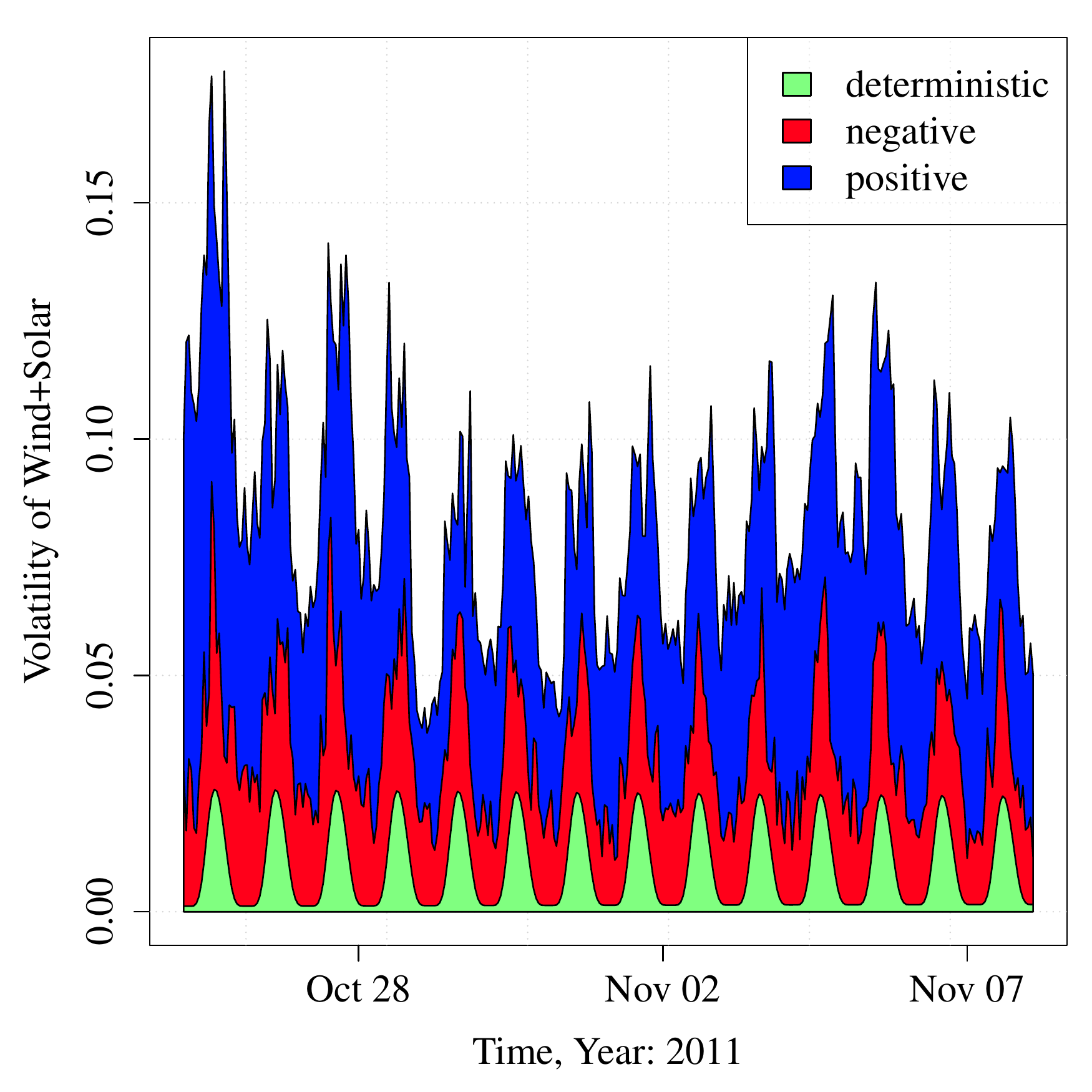}
  \caption{Decomposition of $\what{\sigma}_t^\RR$}
  \label{fig_volatility_sub3}
\end{subfigure}
 \caption{Decomposition of the estimated volatilities $\what{\sigma}_t^i$ into their deterministic seasonal part, the part influenced
 by positive residuals and positive ones for the price, load and wind+solar.}
 \label{fig_volatility}
\end{figure}

Moreover, analyzing the volatility structure determined by $\bssigma_t$ yields promising results. 
Using the TARCH model for $\bseps_t$ allows us to decompose $\bssigma_t$ into three components.
The first one is determined by the periodic coefficient $\bsalpha_0(t)$. It 
describes the deterministic part of the volatility and is automatically a lower bound for $\bssigma_t$.
The second and third components describe the residual impact of the positive and negative residuals, respectively. 
Note that in an ARCH model, both components are equal. The decomposition of the 
estimated volatilities $\what{\sigma}_t^i$ is given in Figure \ref{fig_volatility}.
For illustrative purposes, we selected approximately two weeks of October/November 2011. It can be seen that the volatility of each of the three time series is influenced by a deterministic seasonal component. Moreover, for the price process, we cannot observe a strong leverage effect, the impact of past negative residuals 
is almost equal to the impact of the past positive ones. This is represented by the area marked as \textit{negative} in the left picture being approximately equal to the area marked as \textit{positive}. This relation is obviously different for load time series. Here we observe a strong leverage effect. The impact of negative residuals is higher than the impact of positive residuals. For the wind and solar feed-in an inverse leverage effect is present, as the \textit{positive} area clearly dominates the \textit{negative} area.

Furthermore, we can perform tests to check the significance of the leverage effect. This will be especially interesting for the price series, as judging from our subset of October/November 2011 it was not clear whether a significant effect occurred.
Therefore we evaluate several tests $H_0: A_{i,k} = \sum_{j=1}^k \alpha_j^{+,i} - \alpha_j^{-,i} = 0$ 
for $i \in \II$. The $k$ indicates the lags upto which the leverage effect is taken into account. 

 \begin{figure}[hbt!]
\centering
 \includegraphics[width=0.69\textwidth]{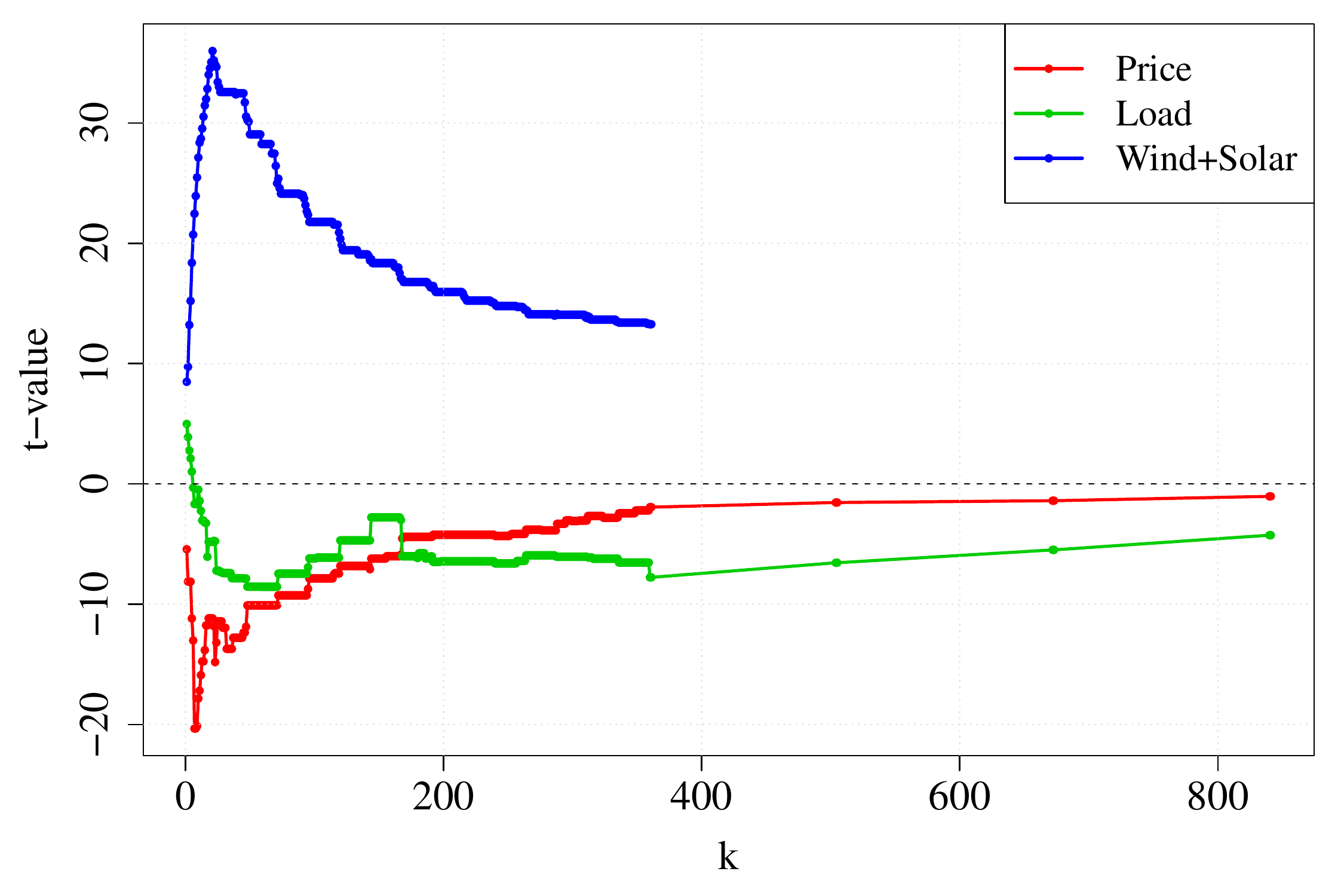}
 \caption{t-values of $A_{i,k}$ for $i\in \II$.}
 \label{fig_levtvals}
\end{figure}

 Figure \ref{fig_levtvals} provides the t-values of the corresponding test for all possible $k$. In order to construct the different values for $k$ we simply cut the remaining model parameters of the TARCH-part after $k$ and performed the test described above. We therefore explicitly state that we did not re-estimate the model for every $k$. 
\par 
It is obvious that the wind and solar energy has clearly positive t-values with its peak at about a day. Those positive t-values indicate a strong inverse leverage effect. 
For the load time series the leverage effect seems to point in the opposite direction. Including all model parameters we can state, that there is a significant standard leverage effect within the data. This is also supported by the t-value for the load in Table \ref{TARCH-t-val}. However, the picture becomes fuzzy when the electricity price is considered. Judging by the Figure it can be obtained, that the electricity prices firstly experiences a positive leverage effect, which diminishes after a certain amount of time. When all parameters up to $k$ are considered, the leverage effect even becomes insignificant, as can be seen in Table \ref{TARCH-t-val}. However, applying a variance model which can account for a leverage effect is still important also for electricity prices, as only those models can account for such a complex structure with different signed impacts of different lags on the volatility.

\begin{table}[ht]
\centering
\begin{tabular}{rcccc} 
  \hline
 & $\what{A}_{i,k=\max}$ & $\what{\sigma}(\what{A}_{i,k=\max})$ & t-value & p-value \\ 
  \hline
Price ($i=\PP$)& -0.0561 & 0.0540 & -1.0394 & 0.2986 \\ 
  Load ($i=\LL$)& -0.1780 & 0.0417 & -4.2666 & 0.0000 \\ 
  Wind+Solar ($i=\RR$)& 0.8427 & 0.0635 & 13.2689 & 0.0000 \\ 
   \hline
\end{tabular}
\caption{Test for the leverage effect $H_0:A_{i, k}=0$, including all estimated parameters.}
\label{TARCH-t-val}
\end{table}

In the same way we can perform an approximate test as to whether an increasing load or wind-solar feed-in leads to 
a significantly increasing or decreasing price in the long run.
These tests have  the null hypothesis $H_0 : \Phi_{\PP,\LL} = \sum_{k\in I_{\PP,\LL}}  \phi^{\PP, \LL}_{k} = 0$ 
and $H_0 : \Phi_{\PP,\RR} = \sum_{k\in I_{\PP,\RR}}  \phi^{\PP, \RR}_{k} = 0$. 
Applying this test to the data (see Table \ref{tab_effects}) it can be concluded that
there is only a small long run impact of the load to the electricity price, which is statistically not significant.
In contrast, an increasing in the amount of wind or solar power feed-in causes a decrease in the price. 
Taking the standardization of $Y_t^i$ by its mean and standard deviation 
before the estimation of \eqref{eq_main_model} into account we can quantify these results.
An 
increase of 1 GWh in the load changes the price by $-0.108 
(\pm 0.499)\footnote{\text{The the term in parentheses gives the symmetric $90\%$ confidence interval}} \frac{\text{EUR}}{\text{MWh}}$
and an increase of 1 GWh wind or solar power feed-in changes the price by $-2.031 ( \pm 0.375) \frac{\text{EUR}}{\text{MWh}}$. 
This result concerning the wind and solar feed-in is consistent with the recent literature on 
quantifying the impact of renewable energy in Germany, see e.g. \citep{wurzburg2013renewable}. 

Moreover, we estimated a linear trend in $Y_t^\PP$ which is negative and significantly different from zero,
so over time, the price seems to decrease. But this effect is not that strong: over one year,
the price changes by $-0.0200 (\pm 0.0069)\frac{\text{EUR}}{\text{MWh}}$. 
Nevertheless, we have to be careful with extrapolating a linear trend far in the future, as it is economically impossible
for such a negative trend to last forever.

\begin{table}[tbh]
\centering 
\begin{tabular}{rrrr|rrrr}
   \multicolumn{4}{c|}{load on price} & \multicolumn{4}{c}{wind+solar on price} \\
  $\what{\Phi}_{\PP,\LL}$ & $\what{\sigma}(\what{\Phi}_{\PP,\LL})$ & t-value & p-value&
  $\what{\Phi}_{\PP,\RR}$ & $\what{\sigma}(\what{\Phi}_{\PP,\RR})$ & t-value & p-value \\ \hline
  -0.064 & 0.180 & -0.356 & 0.721   &-0.732 & 0.082 & -8.921 & 0.000 \\ 
 \hline
\end{tabular}
\caption{Tests for effects of the regressors on the price, two-tailed test.}
\label{tab_effects}

\end{table}

\section{Forecasting} \label{Forecasting}
 Given the estimated model \eqref{eq_main_model} of the sample $(\bsY_1, \ldots, \bsY_n)$ we
 can easily carry out a forecast.
 Let $\bsY_t = g(\bsY_{t-1}, \bsY_{t-2}, \ldots)$ be the representation of  \eqref{eq_main_model}. Then
 we can compute $\what{\bsY}_{n+h}$ iteratively by
 $$\what{\bsY}_{n+h} = g(\what{\bsY}_{n-1+h}, \what{\bsY}_{n-2+h}, \ldots) $$
 and defining $\what{\bsY}_{t}:= \bsY_t$ for $t\leq n$.
 
We performed a forecasting study, where we choose subsequences of the $n=31465$ observations. So we choose 
a data part $H_l$ that is approximately two years long (exactly 18481 observations = 110 weeks + 1 hour),
starting at observation $l$. The $l$ is chosen so that the observed sample ends with an observation
for the electricity price from 23h-24h. 
Finally, we performed a complete estimation on the data sample $H_l$ and estimate the next $h=672$ hours, that is four weeks.
This method provides a fair estimation technique, as we use the same amount of past observations for every forecast.
Given our sample the procedure provides $N=506$ observations, while $\what{\bsY}_{h,k}$ and $\bsY_{h,k}$ for $k \in \{1,\ldots, N\}$ 
denote the corresponding predicted values and observation, respectively. 
Note that we are forecasting the three-dimensional process $\bsY_t$. 
However, in this paper we are interested in the price process $Y_t^\PP$, therefore the further discussion will focus on this process.

The first benchmark we consider is the homoscedastic solution of model \eqref{eq_main_model}.
We get its forecasts automatically by taking the estimated parameters after the first iteration step within the iterative estimation algorithm. Thus, we can directly see the impact of the considered TARCH part. 

 As other own benchmarks we consider the weekly persistent process $Y_t^\PP = Y_{t-168}^\PP$ as persistent model, 
 such as two AIC selected (V)AR processes with time varying mean. 
  Their model is given by
 $$\bsX_t = \bsmu_j \bsone_{\{t \in  (168 \N + j) \}} +  \sum_{k=1}^p ( \boldsymbol{\Phi}_k 
 \bsX_{t-k}- \bsmu_j \bsone_{\{t \in  (168 \N + j) \}}) + \bseps_t$$ 
     where $j \in \{1, \ldots, 168\}$. We consider the univariate choice $\bsX_t = Y_t^\PP$ and the two-dimensional 
     incorporating the load as well, so
     $\bsX_t = (Y_t^\PP,  Y_t^{\LL})$. We also tried to include the wind and solar feed-in but the out-of-sample performance got worse. We estimated the models in a two step approach, first removing the weekly mean and second estimating a
     (V)AR process via Gaussian AIC selection, where the $168$ parameters for the mean are ignored. For the estimation process we 
     solve the Yule--Walker equations that provide a guaranteed stationary solution. As maximal possible order
     for the univariate process, we choose $p_{\max}=1210$ and for the two-dimensional one $p_{\max}=555$.
     The estimation of the processes is very fast and done in a few seconds.
     However, in our empirical results, the order $p$ of the AR process is usually automatically chosen to be about 800, 
     whereas the in the two-dimensional VAR case, it is usually 
     about 400, so both cover the weekly mean and conditional mean behavior.
     
Moreover, we tried to use as many models from the recent literature for a benchmark as possible. But unfortunately 
finding the right competitor is difficult for several reasons. First of all many authors only consider positive
observations or delete some chunks of the data (e.g. outliers, holidays, etc.). Second, some of
them consider information from random regressors such as the load as known, which in a real world situation would not be the case. And finally there are some models, especially from machine learning, 
that simply require an enormous computational time. Furthermore, note that
feed-forward neural networks with one hidden layer and lagged process $\bsY_t$ as input acts very similar to an
AR($p$) process if the chosen lags in the input layer are covered in an AR($p$) process. 
But, as mentioned, neural networks are extremely time consuming, due to the learning phase, which is often based on random selections. 

However, we consider three benchmarks from the literature that have been applied to electricity prices.
First, an ARMA(5,1) model with trend as well as annual, weekly and daily cycles as suggested as one of the best models
in \cite{keles2012comparison}. Note that their daily cycles vary over the seasons of the year.
Moreover, we consider the functional data analysis approach from 
\cite{liebl2013modeling}. But he also modifies the data: he removes outliers and excludes holidays and weekends. 
In fact, he models the prices by estimating the merit order curve
under consideration of the load subtracted by the wind power feed-in. But for our data we noticed that his model gains
better results when only the load is used. Furthermore, in his studies, two functional principal components were
sufficient to model the data well. But we got better results by using three principal components. In the original paper the 
loading coefficients were predicted using a SARIMA model with periodicity five, whereas we use the same model with
a periodicity seven as we are not ignoring the weekends. Moreover, we remark that the usage of the basic model of \cite{liebl2013modeling} as a benchmark to our model is limited, as his approach was tailor-made for the exclusion of special days like weekends and heavy outliers. It may therefore be the case that our reported model performance for his model is only due to the violation of some of his basic assumptions.
As a third benchmark from the existing literature we employ the wavelet-ARIMA approach from  \cite{conejo2005day}. Such a model is often used as a benchmark in electricity spot price 
forecasting. In our application we use the Daubechies 4 wavelet. For modeling the coefficients of the wavelet decomposition we choose 
ARIMA(12,1,1) processes to capture their autoregressive structure.

\begin{figure}[hbt!]
\centering
 \includegraphics[width=0.49\textwidth]{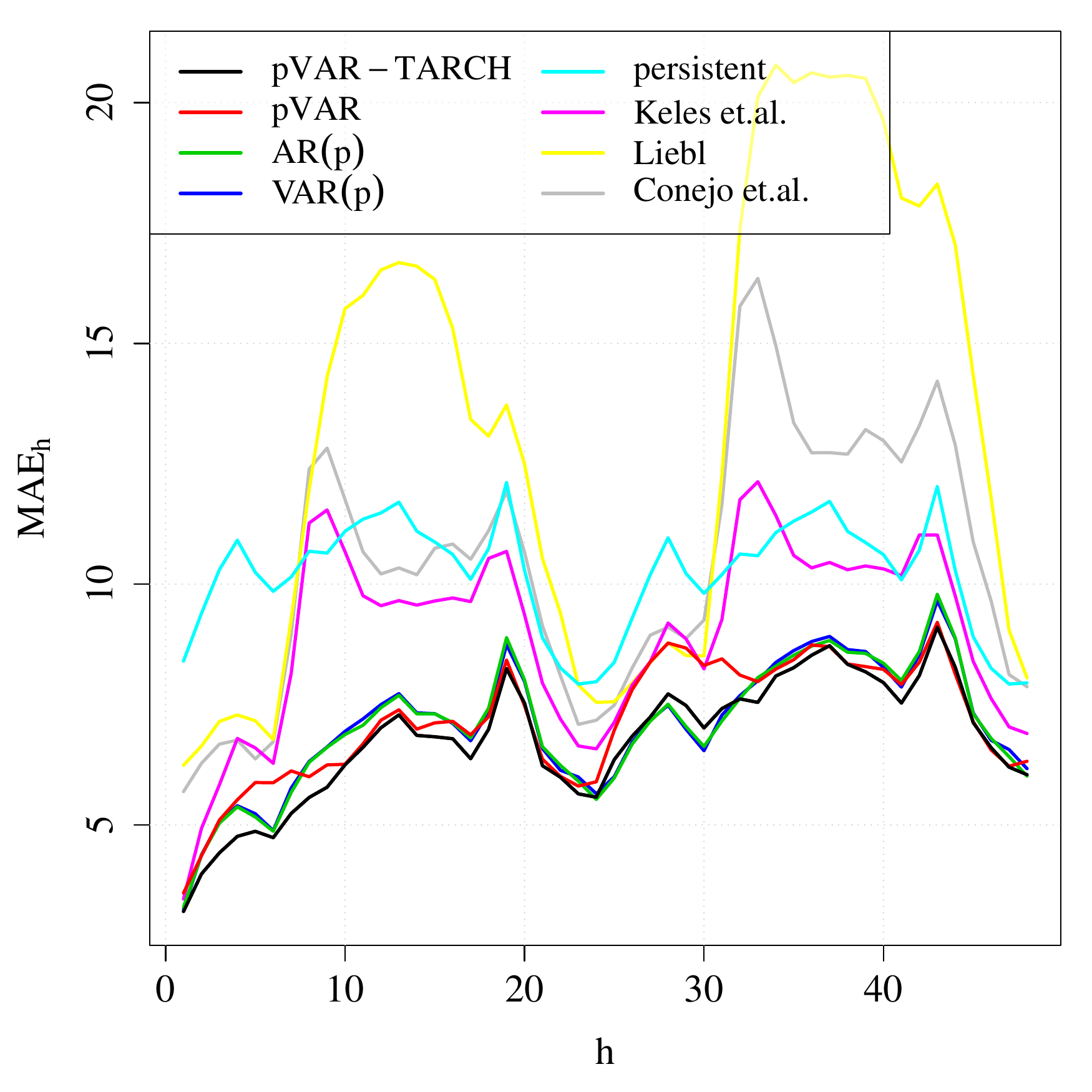}
 \includegraphics[width=0.49\textwidth]{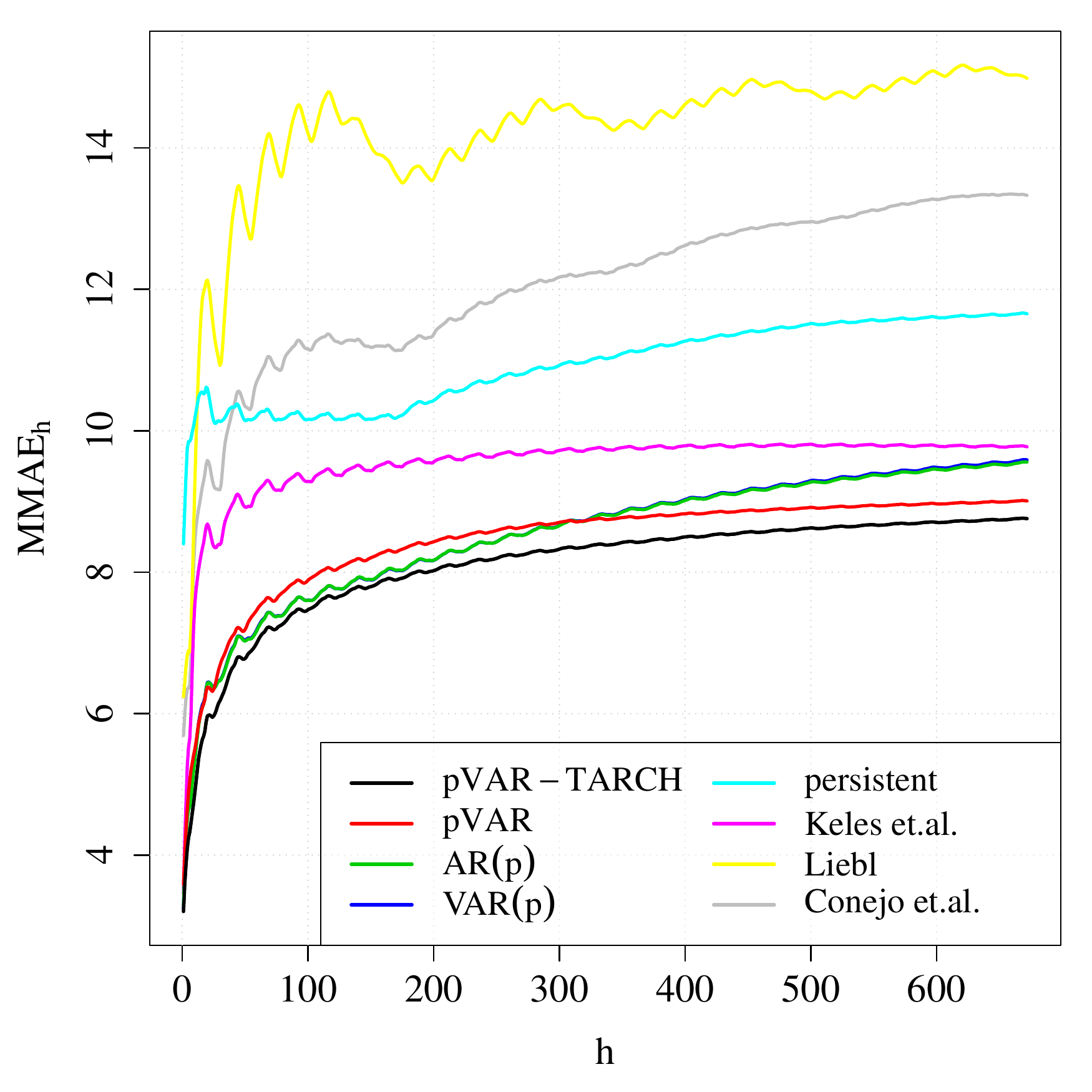}
 \caption{$\MAE^\PP_h$ and $\MMAE^\PP_h$ in EUR/MWh of several models and different forecasting periods. 
 }
 \label{fig_MAE}
\end{figure}

As a performance measure for our forecasting study, we do not use MAPE or WMAPE as is often done in the literature,
because 
the $\text{MAPE}^\PP_h = 
\frac{1}{N} \sum_{k=1}^N \big| \frac{ Y^\PP_{h,k} - \what{Y}^\PP_{h,k} } {Y^\PP_{h,k}} \big| $
is obviously pointless for data that can take zero values.
Instead, we consider the mean absolute forecast error for the prediction time $h$ ($\bsMAE_h$)
as well as the mean of the mean absolute forecast error up to a forecast horizon of $h$ ($\bsMMAE_h$). They are given by
$$ \bsMAE_h = \frac{1}{N} \sum_{k=1}^N |\bsY_{h,k} - \what{\bsY}_{h,k}| \ \ \ \text{ and } \ \ \ 
\bsMMAE_h = \frac{1}{hN} \sum_{j=1}^h \sum_{k=1}^N  |\bsY_{j,k} - \what{\bsY}_{j,k}| .$$
As main criterion to compare different models we suggest the $\bsMMAE_{24}$, as it estimates 
the expected average absolute error for the whole next trading day, which is especially for the practical application of relevance.
The computed $\MAE_h^\PP$ and $\MMAE_h^\PP$ for the considered models 
with selected prediction horizons are given in Table \ref{tab_MAE}. 
The evolution of $\MAE_h^\PP$ and $\MMAE_h^\PP$ with increasing $h$
is presented in Figure \ref{fig_MAE}.

\begin{table}[ht]
\footnotesize
\centering
\begin{tabular}{p{2.7mm}p{5.7mm}p{5.7mm}p{5.7mm}p{5.7mm}p{5.7mm}p{5.7mm}p{5.7mm}p{5.7mm}p{5.7mm}p{5.7mm}p{5.7mm}p{5.7mm}p{5.7mm}p{5.7mm}p{5.7mm}p{5.7mm}}
  \hline
&\multicolumn{2}{p{15mm}}{pVAR-TARCH} &
\multicolumn{2}{c}{pVAR} &
\multicolumn{2}{c}{$\mu$-AR(p)} &
\multicolumn{2}{c}{$\mu$-VAR(p)} &
\multicolumn{2}{c}{persistent} &
\multicolumn{2}{c}{Keles et.al.} &
\multicolumn{2}{c}{Liebl} &
\multicolumn{2}{c}{Conejo et.al.} \\
  \hline 
$h$ & {\tiny $\MAE_h^\PP$} & {\tiny $\MMAE_h^\PP$} & {\tiny $\MAE_h^\PP$} & {\tiny $\MMAE_h^\PP$} &{\tiny $\MAE_h^\PP$} & {\tiny $\MMAE_h^\PP$} &{\tiny $\MAE_h^\PP$} & {\tiny $\MMAE_h^\PP$} &{\tiny $\MAE_h^\PP$} & {\tiny $\MMAE_h^\PP$} &{\tiny $\MAE_h^\PP$} & {\tiny $\MMAE_h^\PP$} &{\tiny $\MAE_h^\PP$} & {\tiny $\MMAE_h^\PP$} &{\tiny $\MAE_h^\PP$} & {\tiny $\MMAE_h^\PP$} 
  \\ \hline
  1 & \textbf{3.20} & \textbf{3.20} & 3.58 & 3.58 & 3.29 & 3.29 & 3.25 & 3.25 & 8.40 & 8.40 & 3.46 & 3.46 & 6.24 & 6.24 & 5.69 & 5.69 \\ 

  4 & \textbf{4.76} & \textbf{4.09} & 5.52 & 4.64 & 5.38 & 4.52 & 5.40 & 4.52 & 10.91 & 9.75 & 6.79 & 5.25 & 7.28 & 6.83 & 6.76 & 6.35 \\ 

  8 & \textbf{5.57} & \textbf{4.60} & 6.00 & 5.31 & 6.30 & 5.01 & 6.32 & 5.04 & 10.68 & 9.99 & 11.27 & 6.66 & 11.96 & 7.81 & 12.39 & 7.48 \\ 

  12 & \textbf{7.02} & \textbf{5.20} & 7.17 & 5.73 & 7.43 & 5.67 & 7.50 & 5.71 & 11.48 & 10.37 & 9.55 & 7.90 & 16.52 & 10.42 & 10.21 & 8.78 \\ 

  16 & \textbf{6.79} & \textbf{5.64} & 7.15 & 6.09 & 7.13 & 6.09 & 7.11 & 6.13 & 10.62 & 10.55 & 9.71 & 8.34 & 15.31 & 11.87 & 10.83 & 9.21 \\ 

  20 & 7.53 & \textbf{5.97} & \textbf{7.48} & 6.37 & 8.01 & 6.43 & 7.98 & 6.44 & 10.29 & 10.60 & 9.37 & 8.68 & 12.50 & 12.13 & 10.66 & 9.58 \\ 

  24 & 5.58 & \textbf{5.95} & 5.90 & 6.31 & \textbf{5.53} & 6.37 & 5.65 & 6.38 & 7.97 & 10.21 & 6.58 & 8.42 & 7.55 & 11.58 & 7.17 & 9.30 \\ 

  168 & \textbf{6.56} & \textbf{7.89} & 7.00 & 8.29 & 6.74 & 8.04 & 6.97 & 8.03 & 7.92 & 10.18 & 7.07 & 9.51 & 7.83 & 13.69 & 8.79 & 11.15 \\ 

  672 & 7.19 & \textbf{8.76} & 7.31 & 9.01 & 8.83 & 9.56 & 8.95 & 9.59 & 8.89 & 11.66 & \textbf{7.00} & 9.77 & 8.82 & 14.99 & 11.22 & 13.33 \\ 
\hline
\end{tabular}
\caption{$\MAE_h^\PP$ and $\MMAE_h^\PP$ in EUR/MWh of several models: 
 pVAR = homoscedastic lasso solution of model \eqref{eq_main_model}, pVAR-TARCH = proposed model.
 Best model is highlighted with bold font. }
  \label{tab_MAE}
\end{table}

\normalsize

It is remarkable that our proposed model can outperform every other one,
especially these that are used in the literature.
Remember that we compare pure out-of-sample methods, that use no information from the prediction time 
during the estimation. Moreover, we noticed that many models are not able to outperform simple AIC selected 
AR($p$) or VAR($p$) models, so e.g. \cite{keles2012comparison} and \cite{liebl2013modeling}. 
Such benchmarks have not not been considered in most of the literature so far, even 
though they are simple and fast to estimate. The relatively good performance is likely
due to the highly carefully chosen model order, so that
hundreds of variables are included in the model, which can cover well the behavior. This amount of
chosen parameters is in the same range as for our reweighted lasso procedure. 
 
 Furthermore, we use Monte Carlo methods to compute the prediction bands conditioned on the given data.
Thus, we can perform a residual based bootstrap on the standardized residuals $\what{\bsZ}_t$ to simulate $\bsY_{t}$.
So we resample from $\what{\bsZ}_t$ to simulate $\Sigma_t$ and afterwards $\bsY_t$ to compute the prediction intervals.

\begin{figure}[hbt!]
\centering
 \includegraphics[width=0.95\textwidth]{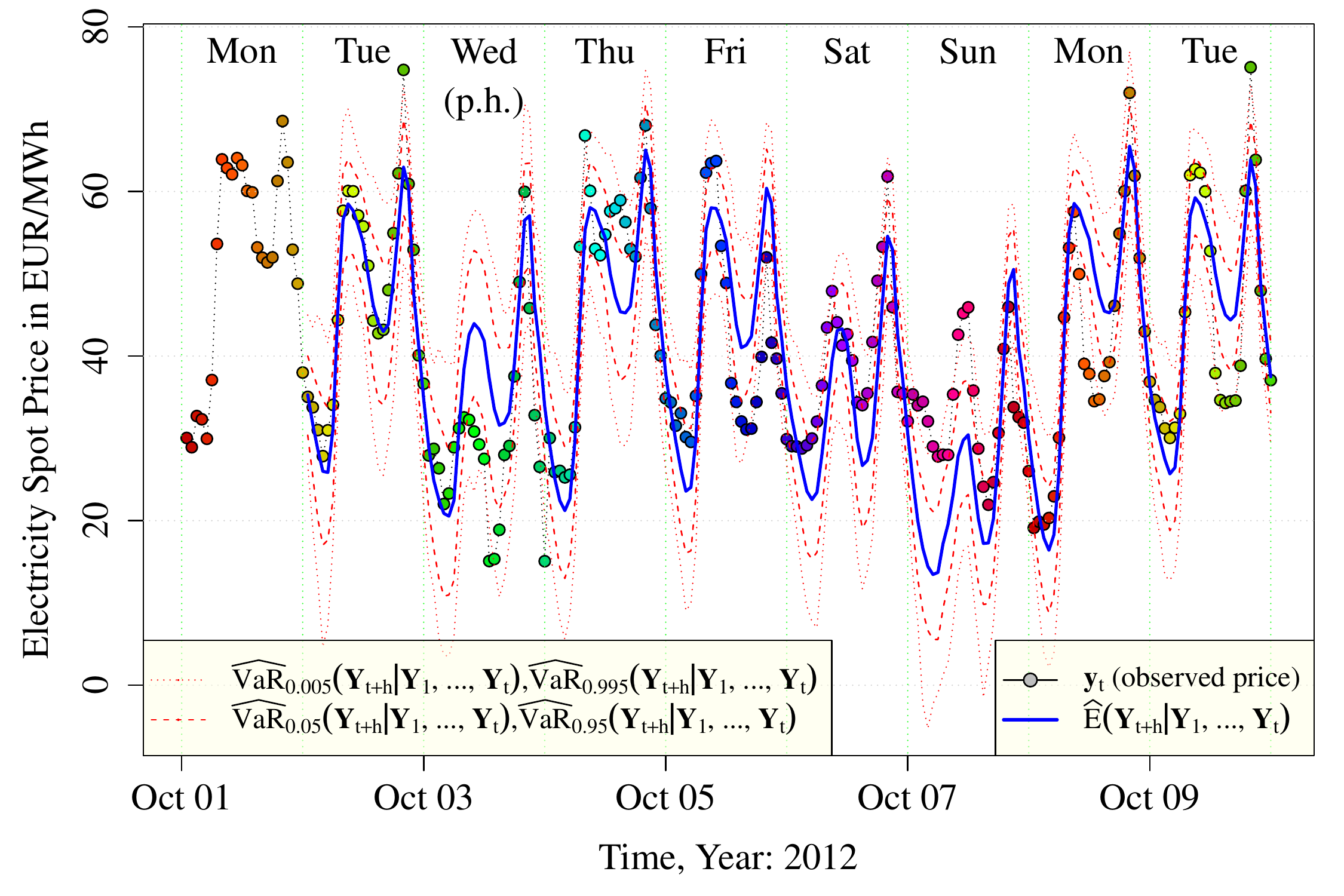}
 \caption{Electricity price prediction with 
   $90\%$ and $99\%$ prediction bands for $h\in \{1, \ldots, 192\}$ using a Monte-Carlo sample size of 1000.
  The 3rd of October is a public holiday (German unity).
 }
 \label{fig_forc}
\end{figure}

For illustrative purposes, we performed a prediction for the sample that is given in Figure \ref{fig_load-wind-solar},
which includes a public holiday. So the considered estimation period is about two years and the prediction horizon is 192 hours.
The corresponding point estimates, such as the estimated conditional mean and the $90\%$ and $99\%$ prediction
intervals are given in Figure \eqref{fig_forc}. The prediction bands are computed by evaluating the symmetric conditional value at risk 
(VaR) resp. quantile levels.

Interestingly, the forecasting performs quite well. For example, Wednesday the 3rd October is an official holiday,
so the forecasted values are significantly 
smaller than those, for example, for the 4th October, even though in this case the forecast still overestimates the real value.

\section{Summary and conclusion} \label{Summary}
The paper presented a model for the hourly electricity price of the European Power Exchange for Germany and Austria.
A periodic VAR-TARCH approach was proposed in order to capture the specific price movements. This model is able to 
deal with the most difficult challenges in modeling electricity prices. Namely, it considers: the mean-reversion 
of prices, the seasonality of the data, the possibility of vast price spikes, negative prices, periodicity, time
dependent variance, the leverage effect, the impact of renewable energy, the impact of the electricity load, 
calender effects like holidays, and time shifts due to daylight saving time. \par
For the necessary simultaneous modeling of the price, load, wind and solar power, we used a modern estimation approach,
which is efficient and turned out to enable rapid estimations. By fitting our model to the before mentioned data 
sets we were able to show insides towards the leverage effect of our considered time series. Moreover, we provided evidence for the effect of renewable energy in reducing the price of electricity. 
Our study showed that an increase in combined wind and solar power of 1 GWh leads to a 
decrease in price of 2.03 $\frac{\text{EUR}}{\text{MWh}}$. \par 
An extensive forecasting study showed that in MAE and MMAE our model outperforms every model which was
used as a benchmark, including recently published models in the literature. For the purpose of forecasting, no 
future knowledge was necessary: every forecast was made exclusively with out-of-sample data. Due to its high
efficiency and fast computing time, our model may be of interest for researchers who want to use it as a benchmark
for their own model as well as for energy companies who need to forecast electricity prices.
\par Nevertheless, there is still much research left for the future. 
For instance, as the markets of some European countries are strongly integrated, measuring the import and export 
feeds may lead to better estimation results and provide guidance for political decisions. \par
It is also debatable whether our model can be reasonably applied to other electricity markets. We notice that the
characteristics of the energy portfolio and working day structure of Germany had significant impacts on the price process. 
But those characteristics are not necessarily possessed by other markets. Therefore, introducing a more comprehensive model may 
be an appropriate future task. \par 
Another direction for future work could also be a more detailed examination of the German energy portfolio, by considering the 
time series of prices for other energy resources, e.g. coal or gas. Detecting and using the cointegration between those time series
may also lead to an improvement in the estimation results. \par
From the statistical point of view the considered model can be extended as well.
So it is possible to include non-linear effects and interactions or change points within the variables. Further
one might allow more parameter to vary periodically, as, for instance, it might be that the TARCH parameters change over time.

\bibliographystyle{apalike}
\bibliography{Bibliothek}
\end{document}